\documentclass[11pt,a4paper]{article}

\usepackage[T1]{fontenc}
\usepackage{newtxtext,newtxmath}
\usepackage[margin=2.5cm]{geometry}
\usepackage{amsmath}
\usepackage{graphicx}
\usepackage{booktabs}
\usepackage{caption}
\usepackage{subcaption}
\usepackage{microtype}
\usepackage[numbers,sort&compress]{natbib}
\usepackage[colorlinks=true,linkcolor=blue,citecolor=blue,urlcolor=blue]{hyperref}

\graphicspath{{Figure/}}

\newcommand{\Ma}{\mathit{Ma}}
\newcommand{\Rex}{\mathit{Re}_x}
\newcommand{\cf}{c_f}
\newcommand{\cp}{C_p}
\newcommand{\cD}{C_D}
\newcommand{\linkedref}[2]{\hyperref[#2]{#1~\ref*{#2}}}
\newcommand{\linkedrefpart}[3]{\hyperref[#2]{#1~\ref*{#2}#3}}

\title{\textbf{Skin friction prediction for attached flows based on two-dimensional inviscid solutions}}

\author{\begin{minipage}{0.94\textwidth}
\centering
Mingkun Xia\textsuperscript{1,2,3},
Shule Zhao\textsuperscript{1,2,3} and
Weiwei Zhang\textsuperscript{1,2,3,*}\\[0.5em]
\small \textsuperscript{1}School of Aeronautics, Northwestern Polytechnical University, Xi'an 710072, China\\
\small \textsuperscript{2}International Joint Institute of Intelligent Fluid Mechanics, Northwestern Polytechnical University, Xi'an 710072, China\\
\small \textsuperscript{3}National Key Laboratory of Aircraft Configuration Design, Xi'an 710072, China\\[0.5em]
\small \textsuperscript{*}Corresponding author: Weiwei Zhang, \texttt{aeroelastic@nwpu.edu.cn}
\end{minipage}}
\date{}

\begin{document}
\maketitle

\begin{abstract}
Boundary layer theory and its analytical methods for skin friction coefficients provide an important basis for aerodynamic analysis. However, classical analytical formulas are mostly limited to flat-plate flows. High-fidelity numerical simulations are not only computationally expensive but also yield predictions that are highly sensitive to physical models, numerical schemes, and grid resolution. To overcome these limitations, symbolic AI opens a new pathway to discover novel laws of complex physical systems from data. Using limited data from surface solutions of the Euler equations (pressure coefficient $\cp$, streamwise Reynolds number $\Rex$, local Mach number $\Ma_x$) and the skin friction coefficient $\cf$ from viscous flows over airfoils, we employ symbolic regression to progressively discover a generalizable, interpretable analytical formula chain for fast skin friction prediction in subsonic and supersonic attached flows. From the perspective of physical mechanisms, the discovered analytical expression chain reveals scaling laws for skin friction at different Mach numbers: the basic form captures the logarithmic decay of skin friction along the streamwise direction in the turbulent boundary layer; the inclusion of a pressure coefficient correction term quantifies the effect of surface pressure variation; and the Mach number correction term evolves with flow regimes, transitioning from the compressibility correction term in subsonic regimes to the thermodynamic effects term in supersonic and hypersonic regimes. This knowledge chain exhibits a unified structure across different Mach numbers, and omitting the correction terms under certain conditions recovers classical theoretical forms, further demonstrating its physical consistency. Validation against typical geometries shows that this analytical formula chain achieves a low average integrated skin friction drag prediction error, with good generalization capability across different freestream conditions and geometric shapes.
\end{abstract}

\noindent\textbf{Keywords:} skin friction, symbolic regression, boundary layer, scaling law, data-driven discovery

\section{Introduction}
\label{sec:introduction}

Low-cost methods for estimating skin friction drag have significant engineering application value in aircraft design, and their accurate prediction directly affects the reliability of aerodynamic performance evaluation and the efficiency of optimization design \citep{anderson2011ebook}. Especially under subsonic cruise conditions, skin friction drag can account for more than 50\% of the total drag; under hypersonic low-Reynolds-number flight conditions, this proportion can even reach 90\%. Therefore, developing efficient and high-precision skin friction prediction methods is of great importance for accurately and rapidly assessing performance metrics such as flight range.

Traditional skin friction acquisition methods are mainly divided into two categories: empirical engineering methods and numerical simulation methods. The estimation of skin friction drag for classical airfoils typically starts from the turbulent flat-plate theory and is corrected using the shape factor \citep{gotten2020airfoil}. Such methods are still widely used in preliminary drag evaluation of wings, but they cannot effectively provide the distribution information of skin friction, which is crucial for local aerodynamic load assessment and fine-grained shape optimization. Engineering algorithms based on boundary layer theory are important means for obtaining boundary layer information \citep{sommer1955free,schlichting1961boundary,sasman1966compressible,Bian1980EngineeringCalculationMethod}, but their computational accuracy for complex geometries is difficult to guarantee. With the development of turbulence theory, some researchers have derived theoretical calculation methods for wall-bounded turbulent skin friction based on different velocity profile assumptions or turbulence models \citep{xiao2020precise,xia2021skin,hu2026reconstructing,zhao2025revisiting,zhao2026generalisation}. Such methods have relatively clear physical mechanisms, but their applicability to complex geometries still needs further improvement. In engineering practice, the primary approach for obtaining skin friction distribution is to solve the Reynolds-averaged Navier--Stokes equations using computational fluid dynamics (CFD) methods \citep{lombardi2000numerical,fidkowski2022coupled}. Although CFD methods can provide solutions with relatively high accuracy, applying them to inverse problems such as optimization design often requires thousands of simulations, making the computational cost orders of magnitude higher than real-time requirements. Moreover, numerical simulation methods suffer from dependencies on computational grid density and quality, numerical schemes, and turbulence models.

To strike a balance between computational accuracy and efficiency, researchers have developed engineering algorithms that couple the Euler equations with integral boundary layer equations. In this class of methods, the Euler equation solution is used as an approximation of the outer edge condition of the boundary layer, and the viscous effects are corrected by solving the integral boundary layer equations, thereby avoiding the direct solution of the full Navier--Stokes equations. \citet{hanlong1991transonic} applied this coupling method to solve the flow around a typical transonic airfoil, and \citet{gaible1993numerical} extended it to the analysis of three-dimensional launch vehicles. Such methods can obtain wall flow information at relatively low computational cost; however, their skin friction prediction still relies on empirical relations for closure within boundary layer theory, making it difficult to guarantee accuracy and robustness under complex conditions such as transonic flow or strong pressure gradients. Nevertheless, these methods reveal an important fact: there exists an intrinsic correlation between the inviscid flow features on the surface (e.g. pressure distribution) provided by the Euler equations and the skin friction distribution within the boundary layer. Moreover, the Euler equation solutions exhibit good consistency, with results from different researchers or numerical methods being mutually verifiable. This implies that if the mapping from inviscid features to skin friction can be learned directly, an efficient and generalizable skin friction prediction method could be achieved. This idea naturally aligns with data-driven modeling.

In recent years, data-driven machine learning methods have opened new avenues for aerodynamic force modeling \citep{brunton2020machine,tang2023some,zhang2025scientometric,zhang2025envisioning}. For aerodynamic force prediction, \citet{sekar2019fast} explored the capability of deep learning methods for flow field prediction across different airfoils. \citet{wang2026goal} proposed a target-oriented feature extraction method that significantly improves the modeling accuracy and generalization ability of data-driven surrogate models by constraining the distance and boundary of hidden feature spaces. For distributed force prediction, \citet{hui2020fast} developed a convolutional neural network-based method for predicting airfoil pressure distributions. \citet{shen2023deep} adopted the Pointnet++ architecture and trained the network on a large dataset of shapes, successfully achieving pressure distribution prediction under hypersonic conditions with substantial geometric variations. However, purely data-driven machine learning approaches, despite their advantages in addressing complex nonlinear problems, suffer from issues such as large sample requirements, weak physical interpretability, and limited extrapolation/generalization capability. To further overcome these limitations, physics-embedded/informed machine learning incorporates first principles into feature design, network architecture, loss functions, etc., effectively enhancing model robustness and generalization ability \citep{karniadakis2021physics}. \citet{ribeiro2023unsteady} combined feature dimensionality reduction with machine learning and used concentrated forces to constrain the loss, achieving prediction of skin friction and pressure coefficients for transonic unsteady airfoils. \citet{shule2024machine,zhao2025euler} proposed a method that embeds the Euler equations into machine learning to simultaneously predict wall pressure and skin friction distributions, and \citet{wang2025ed} extended this idea to aerothermal distributions. Although physics-embedded/informed neural networks can improve generalization, their black-box nature remains unchanged, and the explicit functional relationship between inputs and outputs remains unclear.

Symbolic regression, as a white-box modeling tool \citep{cranmer2023interpretable,udrescu2020ai}, differs from black-box models such as neural networks in that it can automatically search for interpretable expressions that describe the relationships among variables on a given dataset, helping to reveal the underlying physical representations hidden behind the data \citep{wang2023scientific,angelis2023artificial,schmidt2009distilling,langley1981data,Guo2024harness}. This fully embodies the core idea of AI for scientific discovery: using artificial intelligence to automatically discover scientific laws from observational data, compensating for the limitations of traditional manual discovery, and accelerating the transformation from data to theory. In recent years, symbolic regression has shown great potential in areas such as differential equation discovery \citep{brunton2016discovering,beetham2021sparse,ma2024dimensional,chen2025symbolic,wang2025symbolic}, turbulence modeling \citep{weatheritt2016novel,shan2025data,yang2025data}, aerodynamic force modeling \citep{haitao2025scaling,xia2026data}, scaling laws \citep{xia2026data,xie2022data,xia2026hierarchical,zhang2026transformations,bempedelis2025extracting}, and constitutive laws \citep{liu2025symbolic}. In turbulence modeling, \citet{weatheritt2016novel} proposed a novel evolutionary algorithm and applied it to the algebraic modification of the RANS stress--strain relationship, demonstrating the feasibility of symbolic regression in improving the constitutive relations of turbulence models. \citet{shan2025data} applied symbolic regression to adverse-pressure-gradient turbulent boundary layer modeling, directly discovering an explicit pressure gradient correction term for the SA model from high-fidelity data, effectively improving the prediction accuracy of separated flows. \citet{yang2025data} further proposed a three-step discovery framework for turbulence closure models, using symbolic regression to derive a universal inner-layer correction and a specific outer-layer decay function for the mixing-length model, achieving generalization capabilities beyond traditional models in both separated and attached flows. In scaling law discovery, \citet{haitao2025scaling} proposed the scaling function learning method, which uses symbolic regression to reconstruct sparse aerodynamic data correlations across different aircraft geometries, successfully extracting a unified form of aerodynamic scaling laws across geometric shapes. At the methodological level, \citet{xia2026data} developed a progressive discovery framework that injects prior knowledge \citep{xia2026hierarchical} and adopts different symbolic search strategies in stages, significantly improving the efficiency and robustness of mining governing laws in complex flow systems. Collectively, these studies outline the evolution trajectory of symbolic regression from single-point correction to systematic modeling, and from specific working conditions to cross-domain generalization.

Building upon this, this paper proposes a knowledge discovery method for skin friction distribution based on progressive symbolic regression. Based on numerical solutions of the Euler equations and combined with limited RANS-computed skin-friction distribution data, this method progressively mines an analytical expression chain for the skin friction coefficient from the inviscid flow features on the surface. The core contributions of this work are: (1) establishing a white-box knowledge chain from flow field features to skin friction distribution, providing clear physical interpretability for the prediction results; (2) revealing a unified mathematical representation of the skin friction scaling law under different Mach numbers, where the expressions under complex conditions can naturally degenerate into simpler forms, demonstrating their inherent physical consistency; (3) validating the extrapolation generalization capability of the proposed expression chain to conditions and geometries outside the training sample space, providing a new tool that balances accuracy and efficiency for fast evaluation of aerodynamic distributed forces.

The paper is organized as follows. \linkedref{Section}{sec:methods} introduces the process of the progressive symbolic regression method. \linkedref{Section}{sec:results} analyzes the physical interpretation of the discovered expression chain and verifies the effectiveness of the method through typical test cases. \linkedref{Section}{sec:conclusions} provides a systematic summary of the key conclusions of this study.

\section{Methods}
\label{sec:methods}

\begin{figure}
  \centerline{\includegraphics[width=0.9\textwidth]{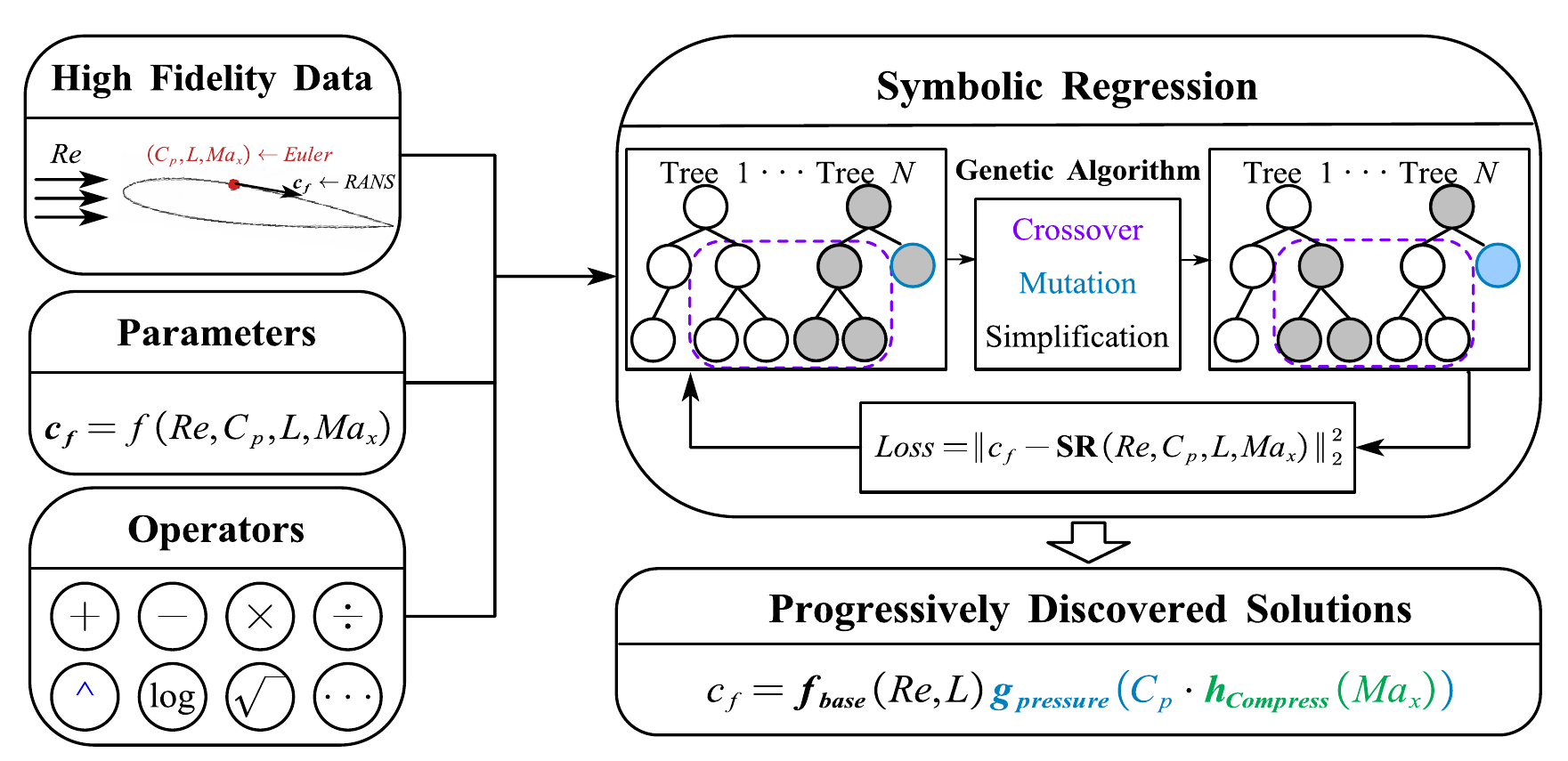}}
  \caption{Workflow diagram of the progressive symbolic regression method. The method is based on numerical simulation data. Guided by a progressive framework that proceeds from simple to complex, the symbolic regression algorithm hierarchically searches and outputs an analytical expression chain that advances layer by layer.}
  \label{fig:workflow}
\end{figure}

This section describes the progressive symbolic regression method adopted, and \linkedref{Figure}{fig:workflow} presents its workflow diagram. The core idea of this method is to decouple the skin friction modeling problem into several physically well-defined sub-problems through a progressive symbolic search strategy that proceeds from simple to complex. Starting from the basic turbulent boundary layer skin friction law, corrections for pressure effects, compressibility effects, etc., are introduced sequentially. At each stage, the expression structure inherited from the previous stage is preserved, and only the correction terms accounting for the newly added physical effects are explored within a restricted search space. This hierarchical progressive discovery framework significantly reduces the search difficulty and ensures the inherent physical consistency of the final expression chain (i.e. the expressions for complex conditions can degenerate progressively into simpler forms). Consequently, a knowledge chain with clear physical interpretability is constructed. The analytical forms of the final expression chain will be presented and discussed in detail in \linkedref{Section}{sec:formula}.

\subsection{Data construction}
\label{sec:data}

The proposed method relies on two types of numerical simulation data: inviscid surface flow features provided by Euler equation solutions, and the corresponding skin friction distribution target values provided by RANS equation solutions. Euler equation solutions are used to obtain inviscid flow features such as surface pressure coefficient $\cp$, local resultant velocity magnitude $V$, local normalized streamwise length $L$, and local Mach number $\Ma_x$. RANS equation solutions, closed with the SA one-equation turbulence model, are used to generate the target skin friction coefficient distributions required for training and validation. All numerical computations are performed using the PhengLEI solver \citep{lei2021design}, whose accuracy and reliability have been thoroughly validated in previous studies. Each test case is computed under turbulent flow conditions. Among them, the first NACA0012 case and the LowRe case adopt the turbulence model proposed by \citet{yang2025data}, while the remaining cases employ the SA turbulence model. For geometry-variation cases, the CST parameterization method is used to describe the airfoil shape: the class function is determined based on the baseline NACA0012 airfoil, the shape function coefficients of the 12-level CST parameters are taken as design variables, and new airfoils are generated by applying perturbations to these coefficients via Latin hypercube sampling (LHS) \citep{stein1987large}. The geometry variation method, freestream condition ranges, and sample usage for each dataset are summarized in \linkedref{Table}{tab:datasets}. The training set is used for expression search in symbolic regression, while the test sets are used to evaluate the generalization capability of the model for unseen operating conditions and geometries.

\begin{table}
\begin{center}
\small
\setlength{\tabcolsep}{2pt}
\begin{tabular}{@{}p{0.36\textwidth}lcccc@{}}
\toprule
Case & Data & $\Ma$ & $Re/10^6$ & $\alpha$ (deg) & For training \\
\midrule
\addlinespace[5pt]
NACA0012 (SR\_Yang) & Train\_1 & 0.15--0.3 & 1--30 & 0--10 & Yes \\
 & Train\_2 & 0.3--0.6 & 1--30 & 0--10 & Yes \\
NACA0012 (SA) & Pre\_incompre & 0.2--0.4 & 3--7 & 0--4 & No \\
 & Pre\_compre & 0.4--0.55 & 7--8 & 4--5 & No \\
 & Pre\_mix & 0.2--0.55 & 3--8 & 0--5 & No \\
Supersonic/hypersonic (NACA0006) & Train\_3 & 1.5/3/6.5 & 1--8 & 0--8 & Yes \\
 & Pre\_super & 2.5/5/8 & 1--8 & 0--8 & No \\
LowRe & Pre\_lowRe & 0.15--0.6 & 0.33--0.4 & 0--4.1 & No \\
New airfoil & Pre\_newairfoil & 0.2--0.5 & 3--8 & 0--6 & No \\
Wing & Pre\_wing & 0.3--0.5 & 3--8 & 0--5 & No \\
\bottomrule
\end{tabular}
\caption{Summary of dataset details. SR\_Yang denotes the general turbulence closure model proposed by \citet{yang2025data}. In the training set, ``Yes'' indicates that the dataset is used for symbolic regression search, and ``No'' indicates that it is used only for generalization performance evaluation.}
\label{tab:datasets}
\end{center}
\end{table}

\subsection{Feature selection}
\label{sec:feature}

The selection of input features significantly influences the prediction accuracy and generalization capability of the model. This paper adopts a strategy combining data-driven screening and physical mechanism analysis to determine the core input variables for symbolic regression. To ensure that the candidate feature library covers the flow information affecting skin friction as comprehensively as possible, this study considers the inviscid flow field characteristics and known freestream parameters. Guided by boundary layer theory, a candidate feature library is constructed, including streamwise coordinate $L$, resultant velocity $V$, freestream parameters (Reynolds number $Re$, Mach number $\Ma$, angle of attack $\alpha$), and local flow parameters (pressure coefficient $\cp$, local Mach number $\Ma_x$). These features characterize the flow state at the boundary layer edge from different perspectives, providing a rich candidate space for subsequent screening.

To identify the features most strongly correlated with skin friction from a data perspective, this paper employs two ensemble learning methods, LightGBM \citep{ke2017lightgbm} and random forest \citep{breiman2001random}, to evaluate feature importance. LightGBM is based on the Boosting strategy, which iteratively trains weak learners and progressively reduces residuals to improve model accuracy. Random forest is based on the Bagging strategy, which uses bootstrap sampling from the original dataset to train multiple decision trees and aggregates their predictions by averaging. The core hyperparameter for both methods is set as $n_{\mathrm{estimators}}=300$, with the remaining parameters at default values. The two algorithms adopt different ensemble strategies (Boosting gradually focuses on difficult-to-fit samples through weighted correction, while Bagging reduces model variance through parallel independent training). Thus, they assess the correlation strength between features and the output target from different perspectives, and the mutual validation of their importance rankings enhances the statistical reliability of the conclusions. The normalized feature importance results are shown in \linkedref{Figure}{fig:feature}. The rankings from the two algorithms exhibit high consistency: streamwise length $L$, pressure coefficient $\cp$, and freestream Reynolds number $Re$ consistently rank as the top three in importance, with relative importance significantly higher than the remaining features. Local Mach number $\Ma_x$, local resultant velocity $V$, and angle of attack $\alpha$ rank in the middle, while freestream Mach number $\Ma$ ranks low. These results indicate that global parameters have a relatively weak direct impact on local skin friction, and their effects are already indirectly reflected through local quantities such as $\cp$ and $\Ma_x$.

Combining physical mechanism analysis with machine learning screening results, and considering both model simplicity and cross-condition generalization capability, this paper ultimately selects four variables -- freestream Reynolds number $Re$, streamwise length $L$, pressure coefficient $\cp$, and local Mach number $\Ma_x$ -- as the core input features for symbolic regression. From the physical essence of boundary layer theory, the dominant factors governing the skin friction coefficient $\cf$ can be summarized into three types of physical effects. The first is the streamwise development effect: $Re$ and $L$ jointly characterize the boundary layer development state and describe the influence of viscous effects on friction drag. The second is the pressure gradient effect: actual airfoil surfaces exhibit favorable or adverse pressure gradients, and the pressure coefficient $\cp$, as a direct measure of the surface pressure distribution in the inviscid flow field, serves as the core inviscid feature representing this effect. The third is the compressibility effect, which becomes increasingly significant as the Mach number rises. The local Mach number $\Ma_x$ is the key parameter capturing this effect.

\begin{figure}
  \centerline{\includegraphics[width=0.75\textwidth]{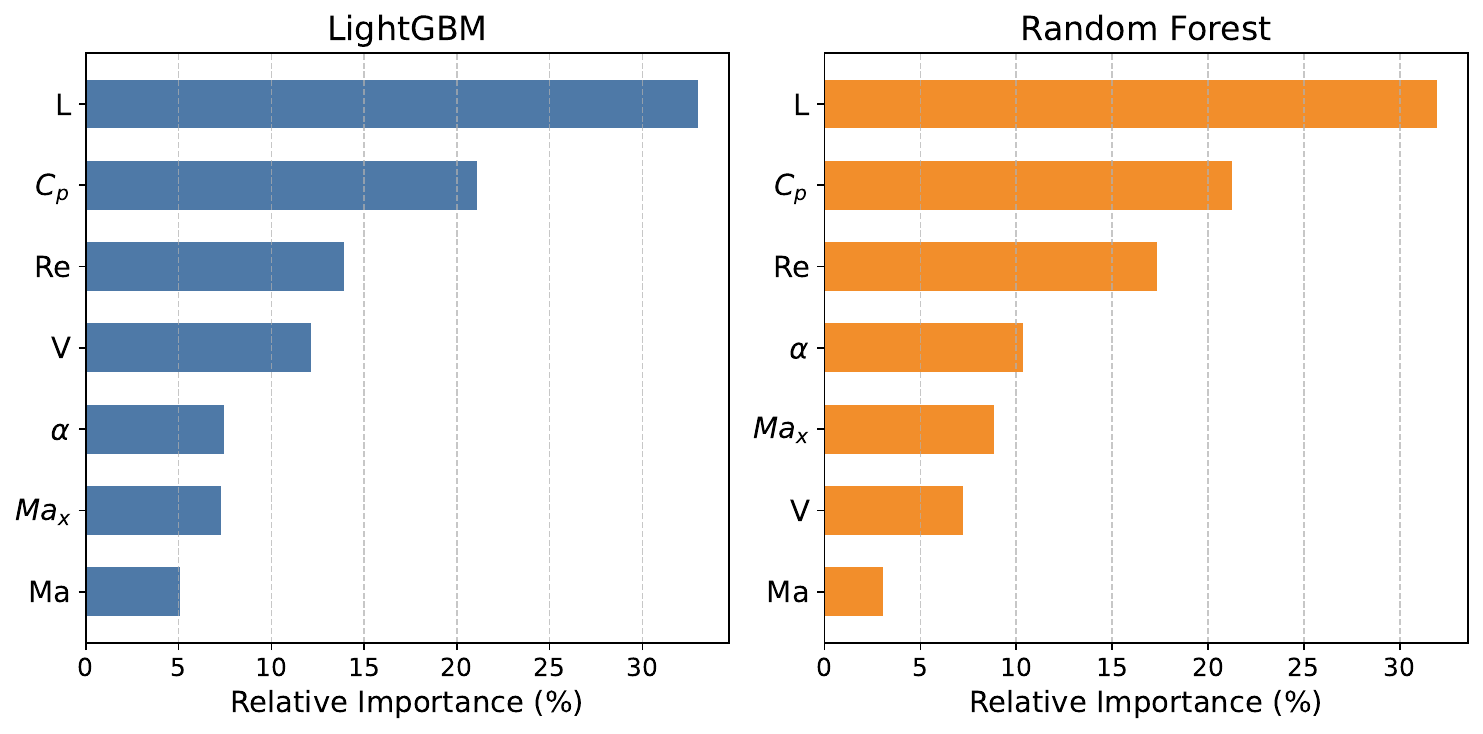}}
  \caption{Feature importance ranking from ensemble learning algorithms. Normalized evaluation results of feature importance from LightGBM and random forest.}
  \label{fig:feature}
\end{figure}

\subsection{Symbolic regression algorithm}
\label{sec:algorithm}

Symbolic regression is a machine learning method that automatically searches for analytical expressions from data. Unlike traditional regression methods that prespecify functional forms (e.g. linear, polynomial), symbolic regression searches the expression space spanned by a given set of operators and variables without assuming a specific functional structure. Therefore, it balances data fitting accuracy with model parsimony, and its results are fully white-box interpretable. This study employs the open-source symbolic regression library PySR \citep{cranmer2023interpretable}, which is based on a multi-population parallel genetic programming algorithm. By simulating natural selection and genetic variation, the algorithm iteratively searches for optimal analytical expressions in the expression space: an initial population of expressions is randomly generated, then in each generation individuals are evaluated according to a fitness function, well-performing expressions are selected as parents, and crossover and mutation operations produce the next generation. Meanwhile, a cross-generational hall-of-fame set is maintained to store the best expressions at different complexity levels. The algorithm runs until a predefined early-stopping condition or the maximum number of iterations is reached.

The hyperparameter configuration of PySR critically affects search efficiency and result quality. Following the principle of balancing search space breadth and model parsimony, this study configures key hyperparameters including the operator set, complexity constraints, penalty terms, and early-stopping conditions. The detailed configuration is listed in \linkedref{Table}{tab:pysr}. The operator set is chosen to include common functional forms encountered in boundary-layer skin-friction scaling laws, such as power laws and logarithmic laws. The upper bound on complexity is set to prevent expressions from becoming too large and losing physical interpretability. The nesting penalty term is used to suppress redundant function nesting (e.g. taking the square root of a squared term), further promoting structural parsimony. The core novelty of this paper is the embedding of the above symbolic regression algorithm into a simple-to-complex, layer-by-layer discovery progressive framework. If all the data were fed into symbolic regression at once, the expression search space would expand exponentially, easily converging to lengthy empirical expressions lacking physical meaning. The detailed process at each stage is described in \linkedref{Section}{sec:formula}.

\begin{table}
\begin{center}
\small
\setlength{\tabcolsep}{4pt}
\begin{tabular}{@{}p{0.18\textwidth}p{0.25\textwidth}p{0.47\textwidth}@{}}
\toprule
Category & Configuration aspect & API setting \\
\midrule
\addlinespace[5pt]
Operator set & Binary operators & $+$, $-$, $*$, $/$, \verb|^| \\
 & Unary operators & \verb|sqrt|, \verb|square|, \verb|cube|, \verb|log| \\
Complexity & Maximum complexity & \verb|maxsize=30| \\
 & Maximum tree depth & \verb|maxdepth=15| \\
 & Constant complexity & 3--5 \\
Penalty term & Symbolic nesting penalty &
\begin{tabular}[t]{@{}l@{}}
\texttt{\detokenize{"sqrt":{"square":0}, "cube":{"cube":0}}}\\
\texttt{\detokenize{"square":{"sqrt":0}, "log":{"log":0}}}
\end{tabular} \\
Other settings & Early-stopping condition &
\begin{tabular}[t]{@{}l@{}}
\texttt{\detokenize{loss < 1e-10 && complexity < 20}}
\end{tabular} \\
 & Loss function & $loss=\|\cf-\mathbf{SR}(Re,L,\cp,\Ma_x)\|_2^2$ \\
\bottomrule
\end{tabular}
\caption{Hyperparameter configuration of the PySR code.}
\label{tab:pysr}
\end{center}
\end{table}

\section{Results}
\label{sec:results}

This section systematically presents the analytical expression chain for skin friction distribution discovered by the progressive symbolic regression, analyzes its physical meaning, and validates its prediction accuracy and generalization capability under different speed regimes and operating conditions.

\subsection{Discovered skin-friction distribution formula}
\label{sec:formula}

Following the progressive symbolic regression framework, this paper successfully discovers a set of analytical expression chains for the skin friction coefficient with clear physical interpretability. The following four discovery stages elaborate on the complete discovery process and hierarchical structure, from fundamental physical laws to multi-effect corrections.

\par\medskip\noindent\textbf{Stage 1: Discovery of the fundamental law.}\par\noindent
Using low-speed incompressible flow over an airfoil as the training data (corresponding to the Train\_1 dataset in \linkedref{Table}{tab:datasets}), the input parameters are the freestream Reynolds number $Re$, streamwise length $L$, and pressure coefficient $\cp$. This stage aims to capture the dominant streamwise development law of turbulent boundary layer skin friction. Through automatic dimensionality reduction, symbolic regression combines $Re$ and $L$ into the local streamwise Reynolds number $\Rex=Re\cdot L$ and discovers the fundamental expression describing the decay of skin friction in the turbulent boundary layer:

\begin{equation}
  \cf^{(0)}(Re,L)=\frac{1}{\{\log(Re\cdot L)\}^{n}},
  \label{eq:stage1}
\end{equation}
where $n$ is a constant to be determined. In general, $n=3.18$ (the symbolic regression result is consistent with the normalized estimate of the classical flat-plate friction formula; see \linkedref{Appendix}{appA} for derivation). This fundamental form is fully consistent with the classical scaling law for turbulent flat-plate boundary layer skin friction (see \eqref{eq:ps}--\eqref{eq:wieghardt} in \linkedref{Appendix}{appA}), verifying the ability of symbolic regression to automatically identify known physical laws from data.

\par\medskip\noindent\textbf{Stage 2: Pressure correction.}\par\noindent
Based on the fundamental expression structure obtained in Stage 1, the form of the fundamental term $1/(\log \Rex)^n$ and the value of the exponent $n$ are fixed, and the search focuses on the correction due to the pressure coefficient $\cp$. Symbolic regression discovers a multiplicative correction term (using the Train\_1 dataset), yielding the extended expression:

\begin{equation}
  \cf^{(1)}(\Rex,\cp)=\cf^{(0)} f_1(\cp)
  =\frac{1-\cp}{(\log \Rex)^n}.
  \label{eq:stage2}
\end{equation}
Here, $1-\cp$ is the pressure correction term. The pressure correction term modulates the friction drag by affecting the boundary layer thickness: a negative $\cp$ (typically associated with flow acceleration) thins the boundary layer, increases the wall velocity gradient, and thus increases the skin friction; a positive $\cp$ (typically associated with flow deceleration) thickens the boundary layer, reduces the wall velocity gradient, and thus decreases the skin friction. Under the flat-plate condition where $\cp=0$, the expression naturally degenerates to the fundamental form of Stage 1. This formula can be derived step by step from boundary layer velocity distribution laws (see \linkedref{Appendix}{appA}) or discovered from data-driven symbolic regression; the two independent approaches corroborate each other, ensuring both data consistency and clear mechanical reasoning.

\par\medskip\noindent\textbf{Stage 3: Subsonic compressibility correction.}\par\noindent
Further fixing the structure of $\cf^{(1)}$, the local Mach number $\Ma_x$ is introduced, and the training is performed using variable-Mach-number data (Train\_2 dataset) to search for the compressibility correction term $f_2(\Ma_x)$. The resulting expression is

\begin{equation}
  \cf^{(2)}(\Rex,\cp,\Ma_x)
  =\cf^{(1)}(\Rex,\cp f_2(\Ma_x))
  =\frac{1-\cp\sqrt{1-\Ma_x^2}}{(\log \Rex)^n}.
  \label{eq:stage3}
\end{equation}
The physical meaning of the Prandtl--Glauert compressibility correction factor $\sqrt{1-\Ma_x^{\smash{2}}}$ is that it scales the pressure coefficient in compressible flow to an equivalent incompressible state, allowing the skin friction formula originally established under the incompressible assumption to be applicable to subsonic compressible flows. Notably, in the expression discovered by symbolic regression, the compressibility correction factor does not appear as an independent multiplicative term but is coupled with the pressure coefficient in a product form ($\cp\sqrt{1-\Ma_x^{\smash{2}}}$). This structure indicates that the Mach number effect does not directly increase or decrease the skin friction but indirectly affects the wall shear stress by modulating the contribution strength of the pressure coefficient. From a physical mechanism perspective, this framework is established for incompressible flow; multiplying by the compressibility factor converts the pressure coefficient in the compressible flow field into an equivalent incompressible pressure coefficient, i.e. $C_{p,\mathrm{inc}}=\cp\sqrt{1-\Ma_x^{\smash{2}}}$, thereby indirectly accounting for compressibility effects within the incompressible skin friction framework. When $\Ma_x\to0$, $\sqrt{1-\Ma_x^{\smash{2}}}\to1$, and the expression naturally degenerates to the Stage-2 form, ensuring physical continuity in the low-speed limit.

\par\medskip\noindent\textbf{Stage 4: Supersonic/hypersonic correction.}\par\noindent
For supersonic/hypersonic conditions (Train\_3 dataset), the structure of $\cf^{(1)}$ is fixed, the specific heat ratio $\gamma$ and the local Mach number $\Ma_x$ are introduced, and the correction term $f_3(\gamma,\Ma_x)$ reflecting thermodynamic state effects is searched. The final expression is
\begin{equation}
  \begin{split}
  &\cf^{(3)}(\Rex,\cp,\Ma_x)=\cf^{(1)}(\Rex,\cp f_3(\gamma,\Ma_x)) \\
  &=\frac{1-\cp(\sqrt{|1-\Ma_x^2|}-\gamma \Ma_x^2\cdot \frac{1}{1+\exp(-0.5(\Ma_x-2.5))})}{(\log \Rex)^n}. 
  \end{split}
  \label{eq:stage4}
\end{equation}
Compared with the subsonic expression,  the Mach-squared term $\gamma {Ma}_x^2$ introduced in \eqref{eq:stage4} reflects the weakening of wall shear stress caused by the boundary-layer temperature rise and density drop due to aerodynamic heating. This weakening is equivalent to adding an extra negative contribution to the equivalent incompressible pressure coefficient. The sigmoid factor $\frac{1}{1+\exp(-0.5({Ma}_x-2.5))} $ in the correction term acts as a smooth activation function, indicating that aerodynamic heating gradually takes effect when the Mach number exceeds approximately 2.5, highlighting the further weakening mechanism dominated by thermal effects (taking $\gamma=1.4$). As the Mach number increases further, this factor converges to 1; for Mach number exceeding 3, it can be replaced by 1, thereby simplifying the numerator to $\sqrt{|1-{Ma}_x^{\smash{2}}|}-\gamma{Ma}_x^{\smash{2}}$. The correction term $\sqrt{|1-{Ma}_x^{\smash{2}}|}-\gamma{Ma}_x^{\smash{2}}\frac{1}{1+\exp(-0.5(\Ma_x-2.5))}$ is always negative in the supersonic regime. Its role is to ensure a positive correlation between the skin friction coefficient $c_f$ and the pressure coefficient $C_p$. The physical reason is that under hypersonic viscous interaction, the growth rate of the boundary-layer displacement thickness $\mathrm{d}\delta^*/\mathrm{d}x$ is proportional to $c_f$, while the local pressure perturbation $C_p$ is proportional to $\mathrm{d}\delta^*/\mathrm{d}x$; hence $c_f$ varies positively with $C_p$. 

\begin{equation}
C_p=\frac{2}{\gamma {Ma}_\infty^2}\left(\frac{P_x}{P_\infty}-1\right)=
\frac{2}{\gamma {Ma}_\infty^2}
\left[
\left(\frac{1+\frac{\gamma-1}{2}{Ma}_\infty^2}
{1+\frac{\gamma-1}{2}{Ma}_x^2}\right)^{\frac{\gamma}{\gamma-1}}-1
\right]
\sim \frac{1}{{Ma}_x^{2\gamma/(\gamma-1)}}.
\end{equation}
In this high-Mach-number limit, the term $\gamma{Ma}_x^{\smash{2}}\frac{1}{1+\exp(-0.5({Ma}_x-2.5))}$ asymptotically approaches $\gamma{Ma}_x^{\smash{2}}$. Since $2\gamma/(\gamma-1)=7$ is larger than the highest order (2) of $\sqrt{|1-{Ma}_x^{\smash{2}}|}-\gamma{Ma}_x^{\smash{2}}$, the term $C_p \cdot f_3(\gamma,{Ma}_x)$ tends to zero as ${Ma}_x\to\infty$, which is consistent with the Mach number independence principle.

\subsection{Validation under multiple conditions}
\label{sec:validation}

To comprehensively evaluate the prediction accuracy and generalization capability of the discovered expression chain, this section conducts validation under typical varying speed regimes, flow conditions, and geometric configurations. Specifically, the validation covers low-speed incompressible conditions, subsonic conditions, supersonic/hypersonic conditions, low-Reynolds-number conditions ($Re\sim10^5$), and variable-geometry conditions. The freestream parameters for each validation case are summarized in \linkedref{Table}{tab:datasets}.

According to the definition in aerodynamics, the total skin friction coefficient $\cD$ can be obtained by integrating the distributed skin friction over the surface. In practical aircraft design, accurate prediction of the total skin friction coefficient is often more critical than the skin friction distribution. Therefore, this paper further defines the relative error of the drag coefficient as the overall accuracy metric:

\begin{equation}
  RE_{f}=
  \left|
  \frac{\oint c_{f,\mathrm{pred}}(s)\,\mathrm{d}s-\oint c_{f,\mathrm{true}}(s)\,\mathrm{d}s}
       {\oint c_{f,\mathrm{true}}(s)\,\mathrm{d}s}
  \right|.
  \label{eq:error}
\end{equation}
In the following case validations, the relative error of the drag coefficient ($RE_{f}$) will be used as the primary metric for evaluating prediction accuracy.

\subsubsection{Low-speed incompressible/subsonic formula validation}
\label{sec:lowspeed}

\begin{figure}
\begin{center}
\begin{tabular}{@{}cc@{}}
\multicolumn{1}{@{}l}{(a) $\Ma=0.29, Re=2.9\times10^7, \alpha=9.2^\circ$} &
\multicolumn{1}{l@{}}{$\Ma=0.29, Re=3.1\times10^6, \alpha=2.36^\circ$} \\
\includegraphics[width=0.47\textwidth]{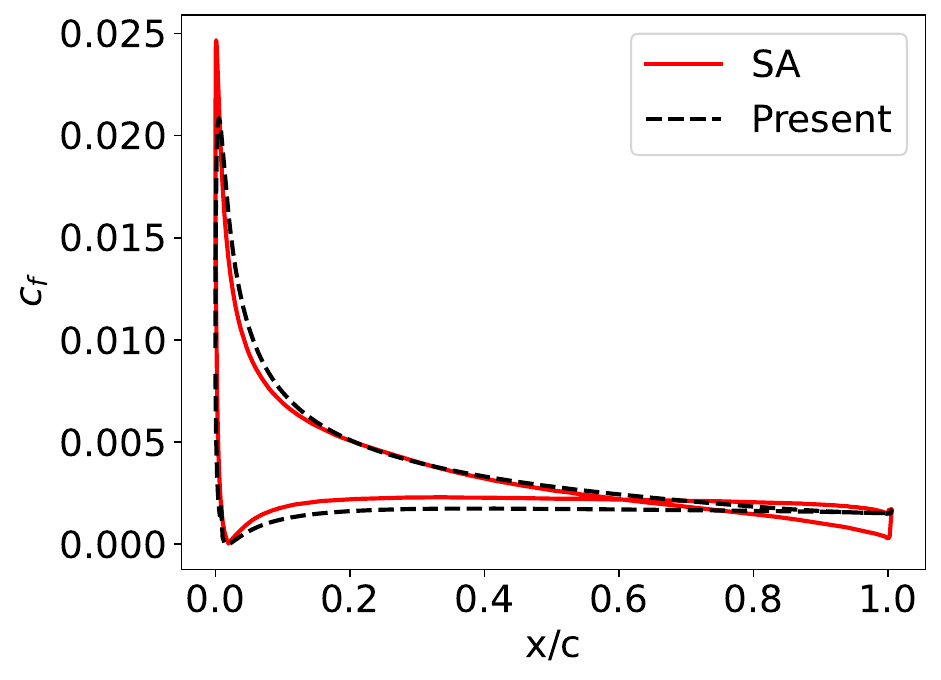} &
\includegraphics[width=0.47\textwidth]{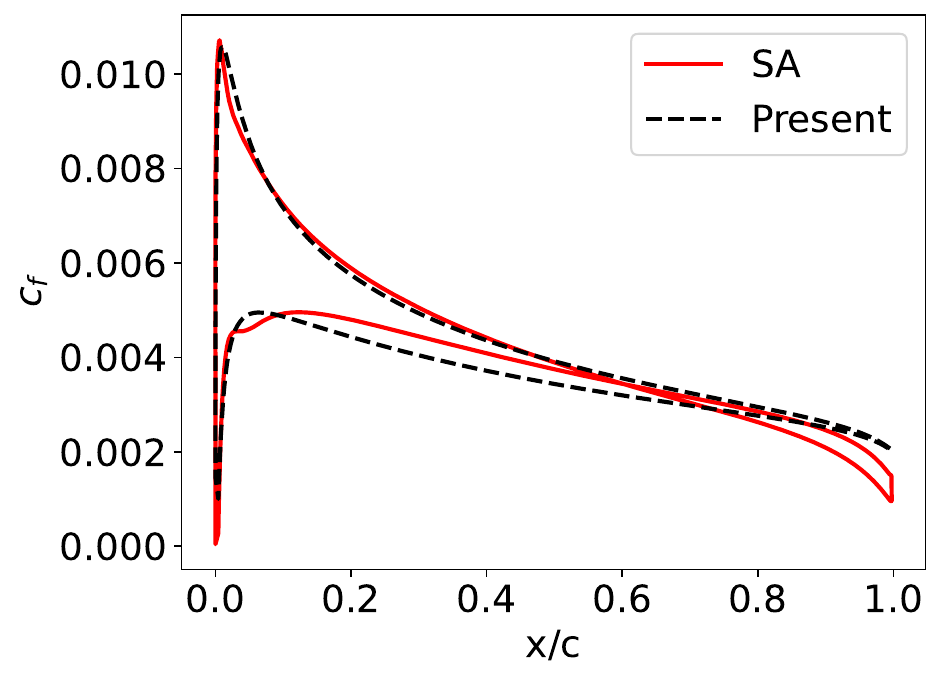} \\
\multicolumn{1}{@{}l}{(b) $\Ma=0.35, Re=3.8\times10^6, \alpha=3.25^\circ$} &
\multicolumn{1}{l@{}}{$\Ma=0.6, Re=1.8\times10^7, \alpha=2.0^\circ$} \\
\includegraphics[width=0.47\textwidth]{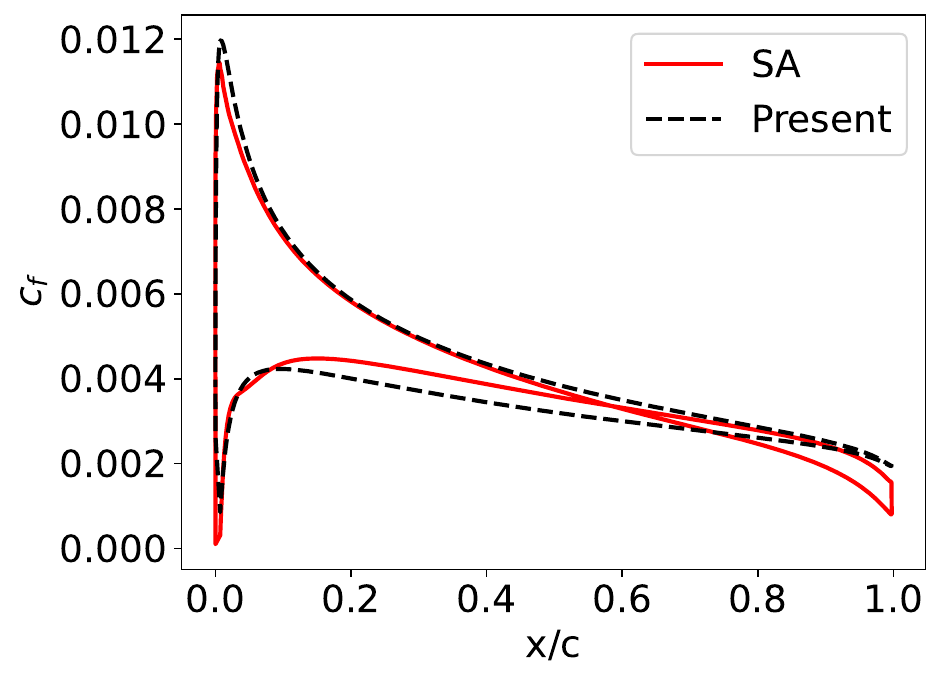} &
\includegraphics[width=0.47\textwidth]{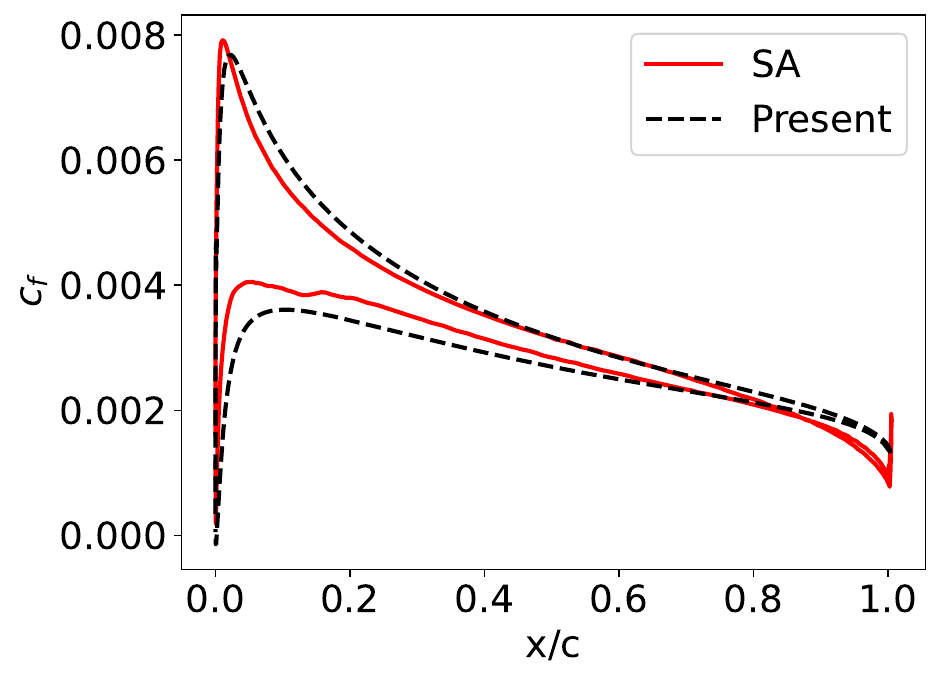} \\
\multicolumn{2}{@{}l@{}}{(c)} \\
\multicolumn{2}{c}{\includegraphics[width=0.72\textwidth]{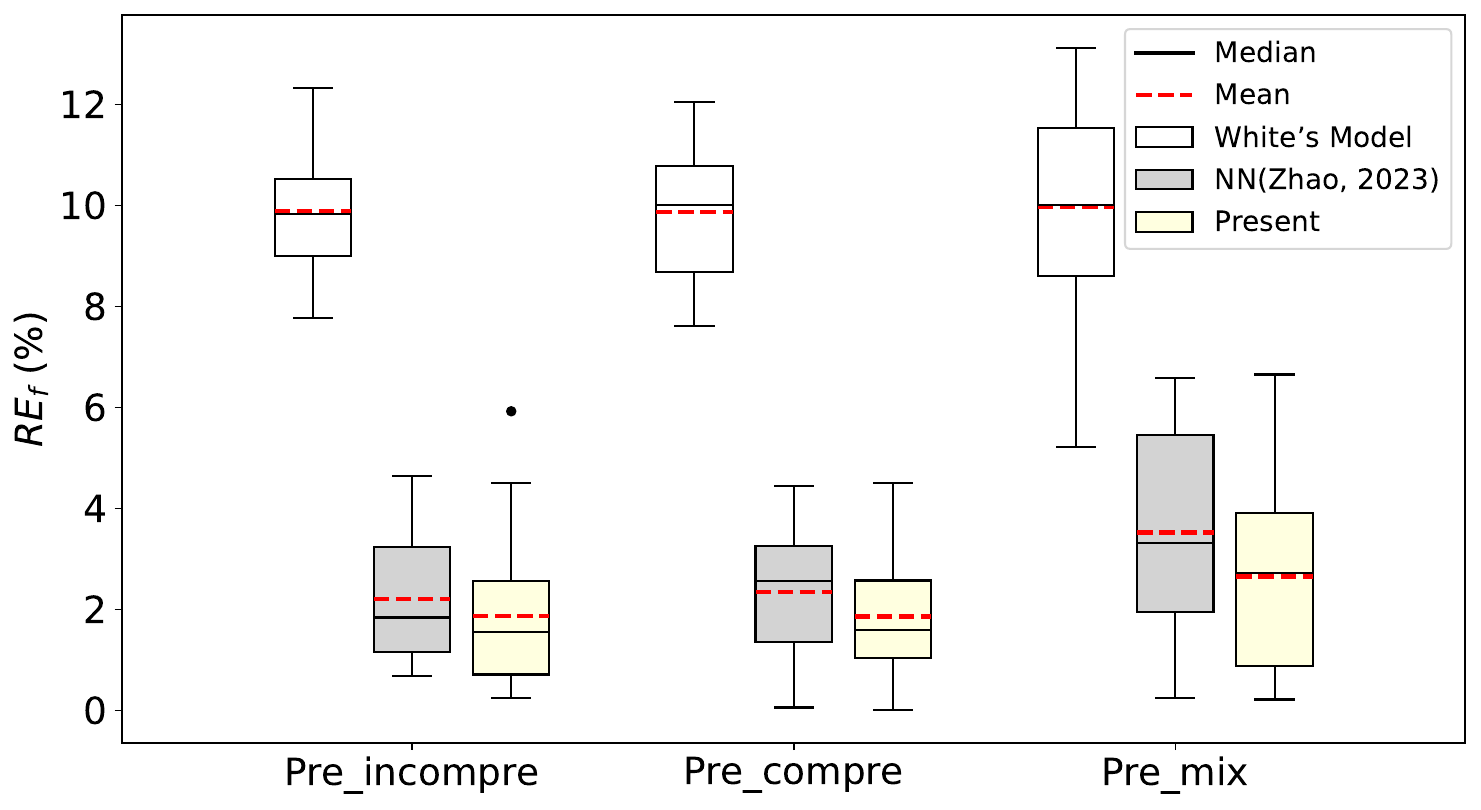}}
\end{tabular}
\caption{Comparison of skin friction distribution predictions for low-speed and subsonic conditions. (a) Validation cases from the Train\_1 and Pre\_incompre datasets; (b) validation cases from the Train\_2 and Pre\_compre datasets; (c) comparison of skin friction predictions among the symbolic regression formula, the classical White formula, and the neural network \citep{shule2024machine}.}
\label{fig:low_subsonic}
\end{center}
\end{figure}

Under low-speed incompressible conditions, compressibility effects are negligible, and the expression given by \eqref{eq:stage2} is adopted. For this speed regime, the Train\_1 dataset from \linkedref{Table}{tab:datasets} is used for model discovery, and the Pre\_incompre dataset is used for validation. \linkedrefpart{Figure}{fig:low_subsonic}{(a)} shows a comparison between the predicted skin friction distribution under typical conditions and the RANS reference solution. The results indicate that the expression accurately captures the skin friction distribution along the airfoil surface for low-speed attached flows.

Under subsonic conditions, the compressibility correction term is activated, and the expression given by \eqref{eq:stage3} is used. For this speed regime, the Train\_2 dataset from \linkedref{Table}{tab:datasets} is used for model discovery, and the Pre\_compre dataset is used for validation. \linkedrefpart{Figure}{fig:low_subsonic}{(b)} compares the predicted skin friction distribution under typical conditions with the RANS reference solution. It can be seen that after introducing the compressibility correction, the model maintains good prediction accuracy under subsonic conditions, and the coupling effect between pressure effects and Mach number effects is effectively captured.

To quantitatively evaluate the prediction accuracy of the discovered expression chain, error statistics are performed on all samples of the three datasets, Pre\_incompre, Pre\_compre and Pre\_mix, and box plots of the integrated drag error are drawn. Furthermore, to examine the accuracy differences among the symbolic regression white-box model, the classical local skin friction coefficient White formula $\cf=0.445/(\ln (0.06\Rex))^2$ \citep{white2011fluid}, and a neural network model that predicts skin friction based on Euler equation solutions \citep{shule2024machine} (where the neural network uses a residual neural network with 15 layers and 32 neurons per layer as the baseline), the comparison of their prediction performance is shown in \linkedrefpart{Figure}{fig:low_subsonic}{(c)}. After statistical error analysis of the skin friction predictions, the white-box expression obtained by symbolic regression yields a drag integration error for skin friction distribution that is comparable to, or slightly better than, that of the black-box neural network model, while also featuring a concise analytical form and clear physical meaning.

\subsubsection{Supersonic/hypersonic formula validation}
\label{sec:supersonic}

\begin{figure}[!t]
\begin{center}
\begin{tabular}{@{}cc@{}}
\multicolumn{1}{@{}l}{(a) $\Ma=1.5, Re=6.0\times10^6, \alpha=0^\circ$} &
\multicolumn{1}{l@{}}{(b) $\Ma=2.5, Re=6.0\times10^6, \alpha=0^\circ$} \\
\includegraphics[width=0.47\textwidth]{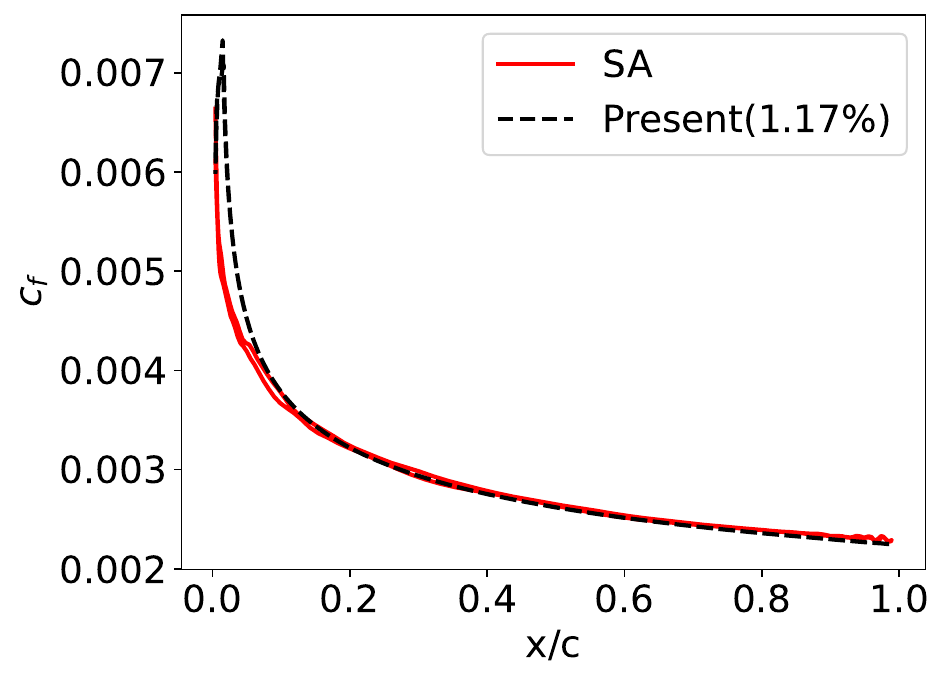} &
\includegraphics[width=0.47\textwidth]{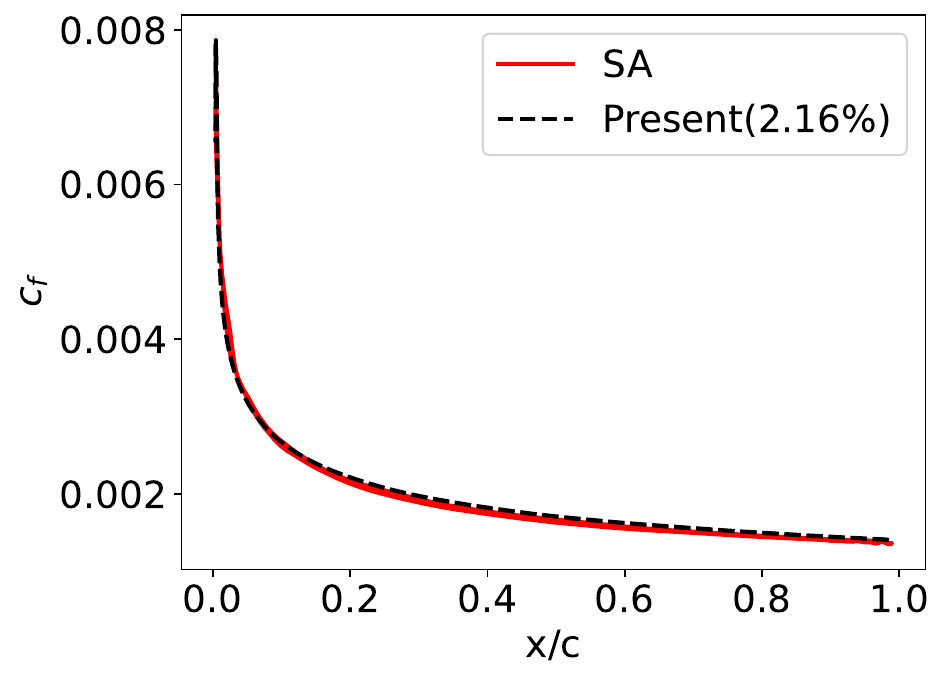} \\
\multicolumn{1}{@{}l}{(c) $\Ma=3.0, Re=6.0\times10^6, \alpha=0^\circ$} &
\multicolumn{1}{l@{}}{(d) $\Ma=5.0, Re=8.0\times10^6, \alpha=2.0^\circ$} \\
\includegraphics[width=0.47\textwidth]{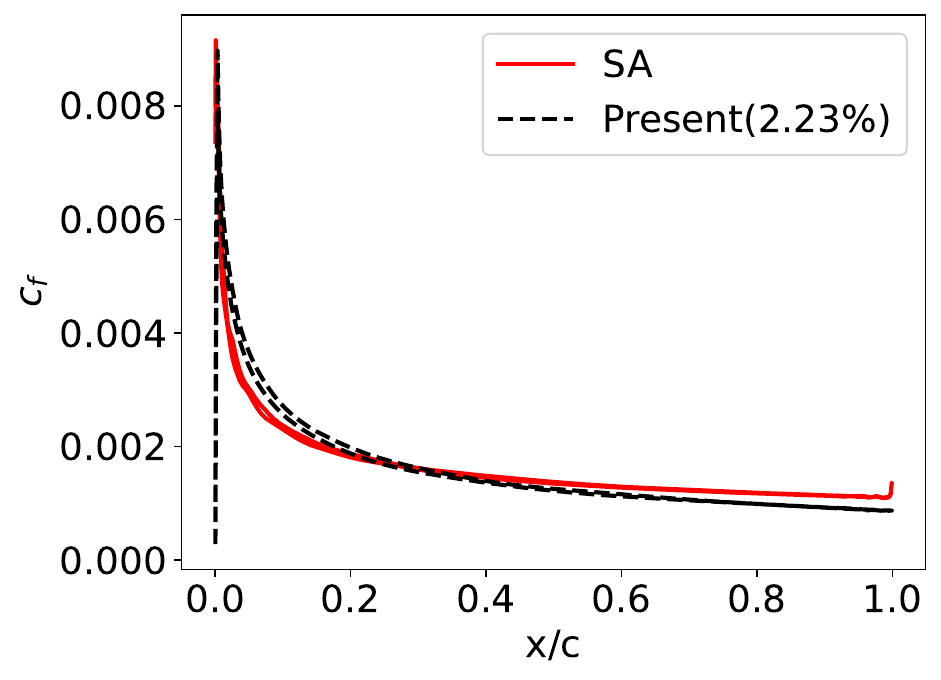} &
\includegraphics[width=0.47\textwidth]{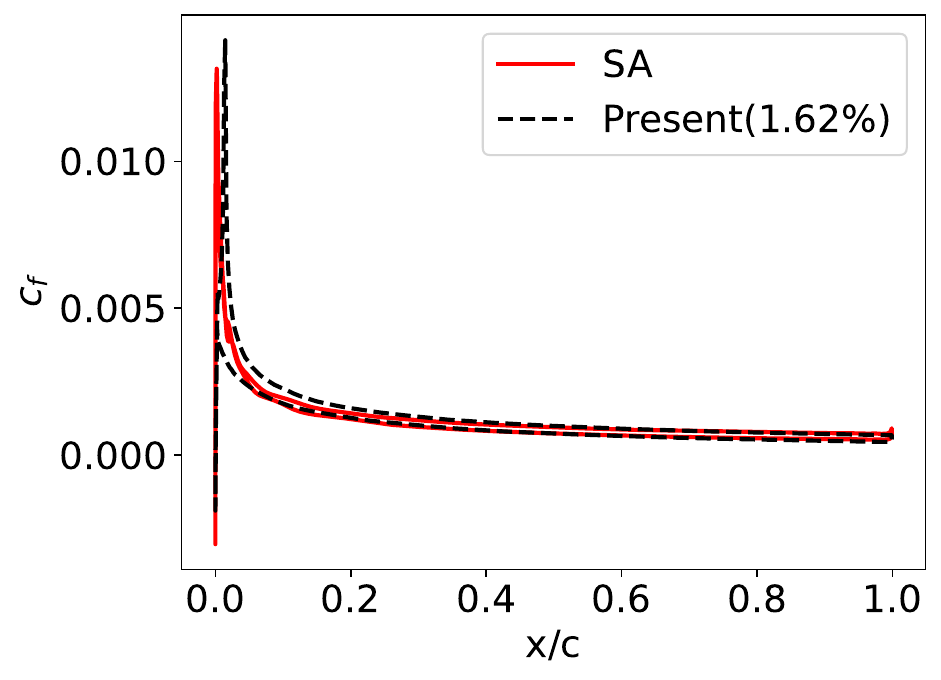} \\
\multicolumn{1}{@{}l}{(e) $\Ma=6.5, Re=1.0\times10^6, \alpha=8.0^\circ$} &
\multicolumn{1}{l@{}}{(f) $\Ma=8.0, Re=8.0\times10^6, \alpha=2.0^\circ$} \\
\includegraphics[width=0.47\textwidth]{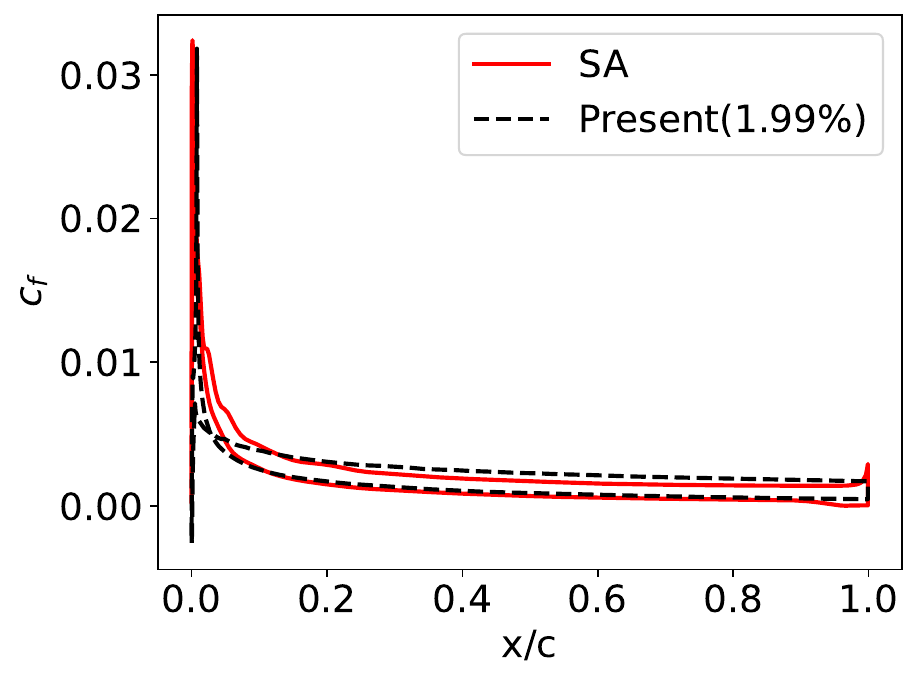} &
\includegraphics[width=0.47\textwidth]{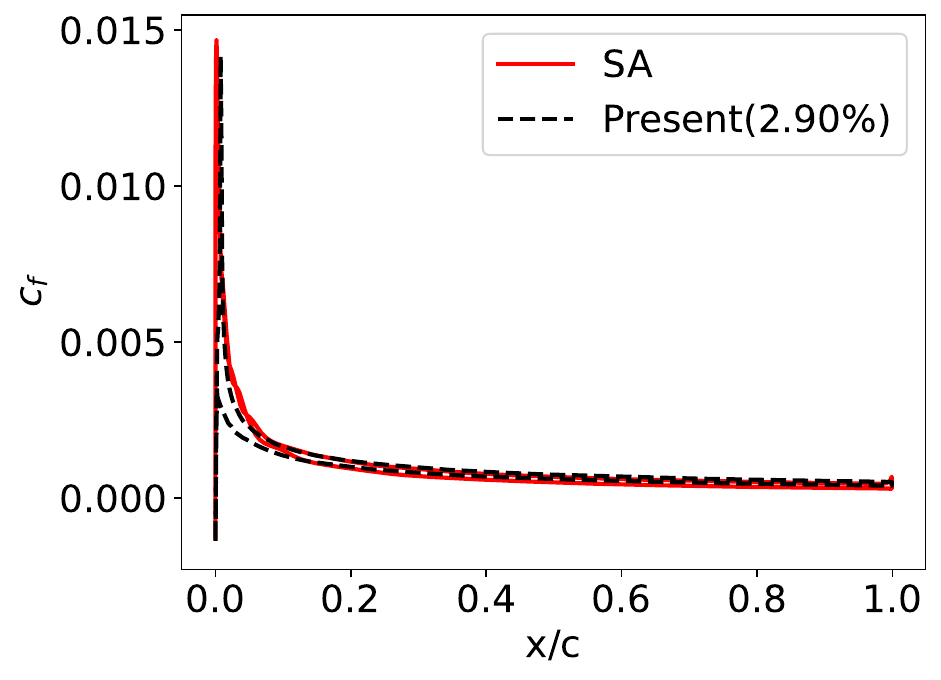}
\end{tabular}
\caption{Validation of the supersonic formula for skin friction distribution prediction under supersonic/hypersonic conditions.}
\label{fig:supersonic}
\end{center}
\end{figure}

Under supersonic/hypersonic conditions, thermal effects become significant. The temperature inside the boundary layer rises sharply and the density drops substantially, leading to severe distortion of the velocity profile and consequent attenuation of the wall shear stress. Under such conditions, the compressibility correction term $\sqrt{1-\Ma_x^{\smash{2}}}$ valid for subsonic flows is no longer sufficient to capture the strong thermodynamic effects. Therefore, the Mach-squared term $\gamma\Ma_x^2$ is introduced as a correction. \linkedref{Figure}{fig:supersonic} shows a comparison between the predicted skin friction distribution under hypersonic conditions and the RANS reference solution. The results indicate that after introducing the thermodynamic correction term, the model exhibits a good capability to capture the trend of skin friction, with consistently low concentrated‑force errors and excellent overall fitting performance across the entire Mach number range from 1.5 to 8.

\subsubsection{Lower-Reynolds-number condition validation}
\label{sec:lowre}

\begin{figure}[!t]
\begin{center}
\begin{tabular}{@{}cc@{}}
\multicolumn{1}{@{}l}{(a) $\Ma=0.15, Re=4\times10^5, \alpha=0^\circ$} &
\multicolumn{1}{l@{}}{(b) $\Ma=0.32, Re=3.5\times10^5, \alpha=0.83^\circ$} \\
\includegraphics[width=0.47\textwidth]{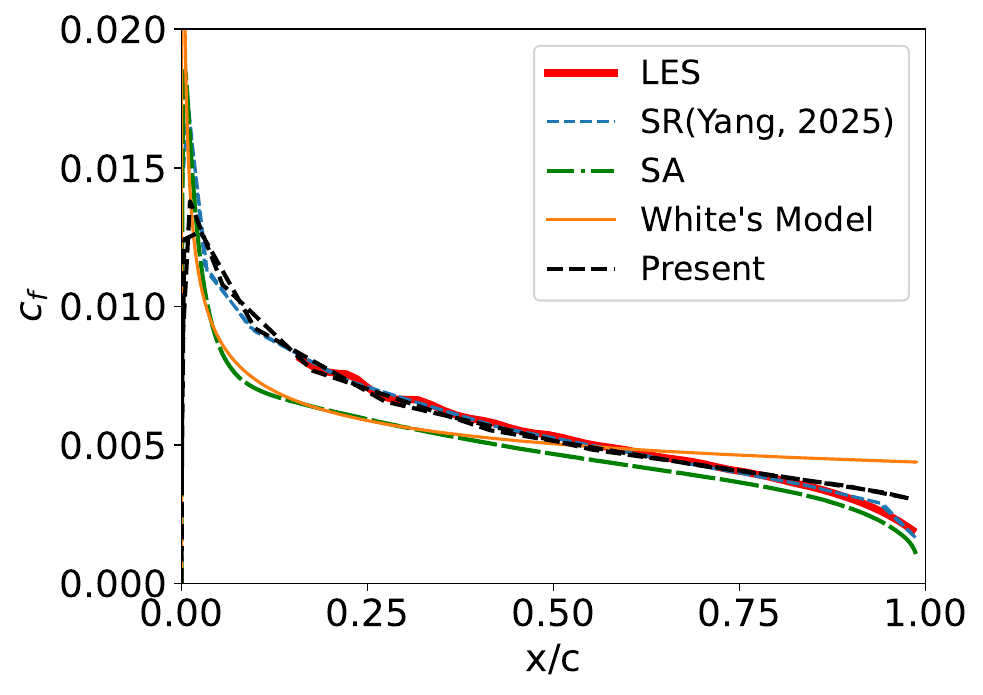} &
\includegraphics[width=0.47\textwidth]{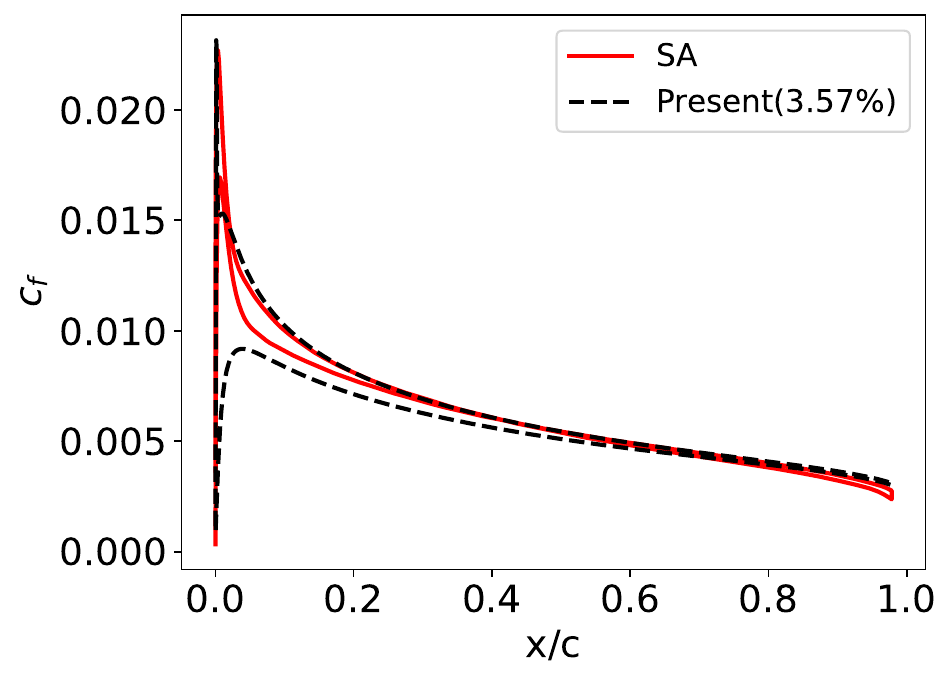} \\
\multicolumn{1}{@{}l}{(c) $\Ma=0.4, Re=3.3\times10^5, \alpha=4.1^\circ$} &
\multicolumn{1}{l@{}}{(d) $\Ma=0.6, Re=3.3\times10^5, \alpha=0.84^\circ$} \\
\includegraphics[width=0.47\textwidth]{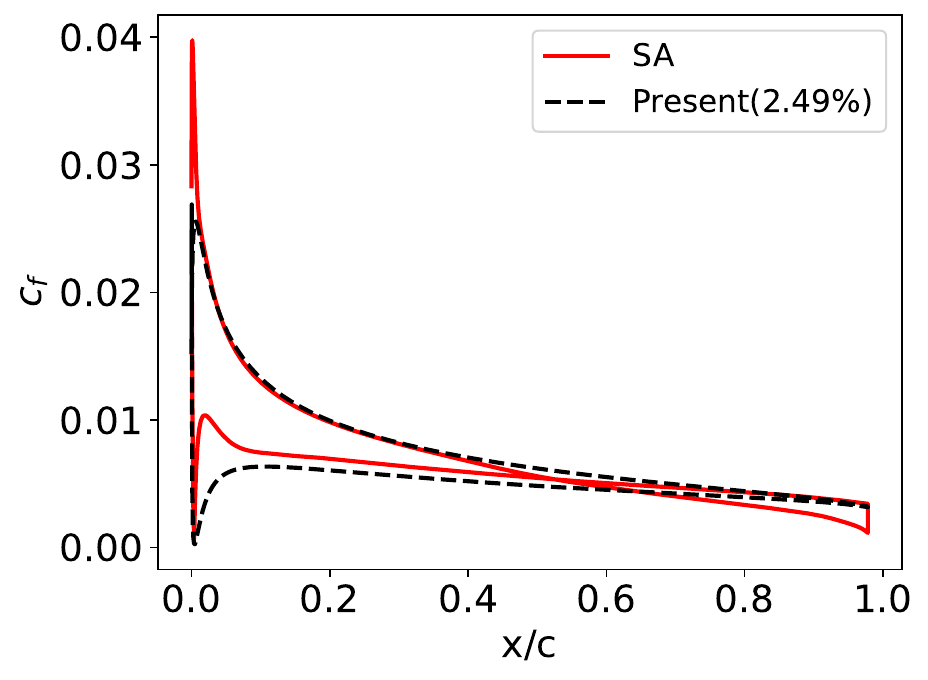} &
\includegraphics[width=0.47\textwidth]{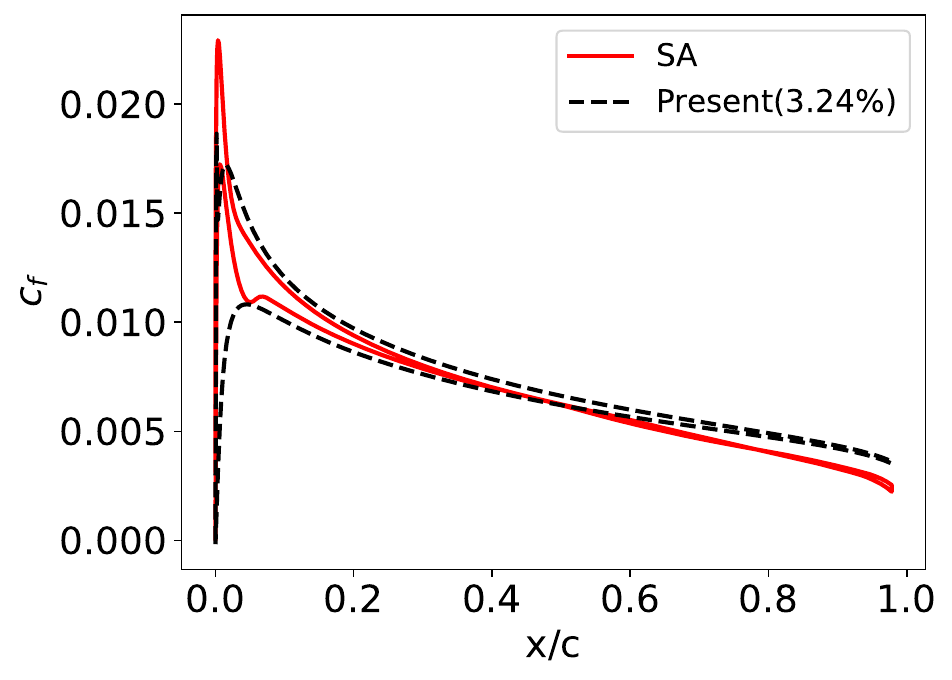}
\end{tabular}
\caption{Validation under low-Reynolds-number conditions (on the order of $10^5$). (a) NACA0012, angle of attack $=0^\circ$, compared with LES data; (b--d) NACA0012, non-zero angles of attack, together with the relative errors of the integrated skin friction drag.}
\label{fig:lowre}
\end{center}
\end{figure}

To examine the applicability of the discovered expression chain under low-Reynolds-number conditions (on the order of $10^5$), the Pre\_lowRe dataset from \linkedref{Table}{tab:datasets} is used for validation. This dataset contains four cases covering different combinations of Mach number and angle of attack. At low Reynolds numbers ($Re\sim10^5$), the relative thickness of the boundary layer increases and viscous effects become more significant. Under these conditions, since the Reynolds number deviates significantly from the training range ($Re\sim10^6$ and $10^7$), the exponent $n$ in the base expression \eqref{eq:stage3} needs to be fine-tuned using a small number of samples, yielding $n\approx3.28$. \linkedref{Figure}{fig:lowre} shows a comparison between the predicted skin friction distributions for the four cases and the RANS reference solutions (\linkedrefpart{Figure}{fig:lowre}{a} also includes LES data). The results indicate that under zero angle-of-attack conditions (\linkedrefpart{Figure}{fig:lowre}{a}), the proposed expression agrees well with the LES data over the middle and aft portions of the airfoil; only a certain deviation is observed near the leading-edge acceleration region, which has little effect on the overall friction drag estimation, with an integrated force error of 0.54\%. In stark contrast, the classical White formula significantly overestimates the skin friction over the entire airfoil surface, exhibiting large deviations from the LES data. Under non-zero angle-of-attack conditions (\linkedrefpart{Figures}{fig:lowre}{b--d}), the predicted skin friction distributions follow the same trend as the RANS reference solutions except for the deviation at the leading edge. This demonstrates the rapid adaptability of the white-box model to extreme conditions.

\subsubsection{Variable geometry generalization validation}
\label{sec:geometry}

\begin{figure}[!t]
\begin{center}
\begin{tabular}{@{}cc@{}}
\multicolumn{1}{@{}l}{(a)} &
\multicolumn{1}{l@{}}{(b) $\Ma=0.44, Re=3.98\times10^6, \alpha=0.503^\circ$} \\
\includegraphics[width=0.47\textwidth]{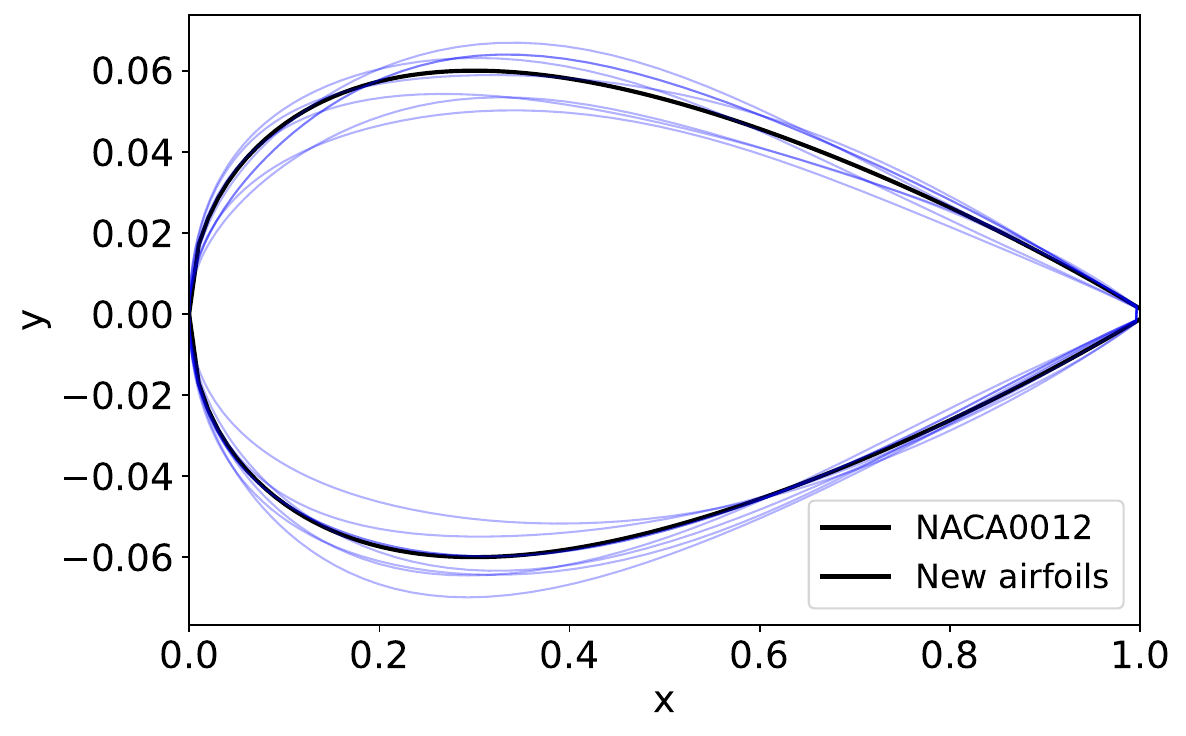} &
\includegraphics[width=0.47\textwidth]{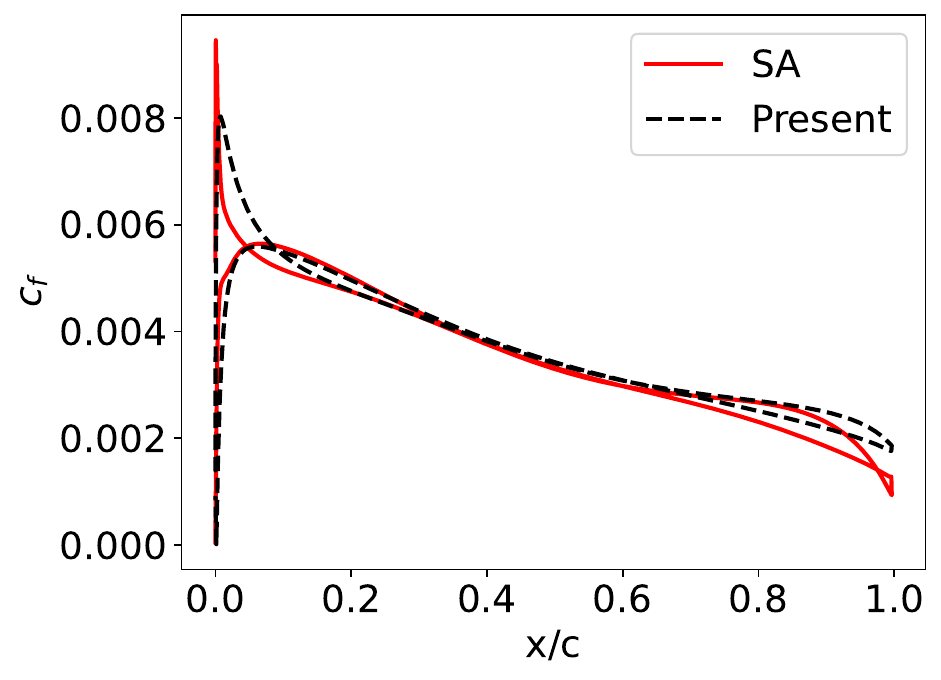} \\
\multicolumn{1}{@{}l}{(c) $\Ma=0.34, Re=3.56\times10^6, \alpha=1.74^\circ$} &
\multicolumn{1}{l@{}}{(d) $\Ma=0.35, Re=5.54\times10^6, \alpha=2.16^\circ$} \\
\includegraphics[width=0.47\textwidth]{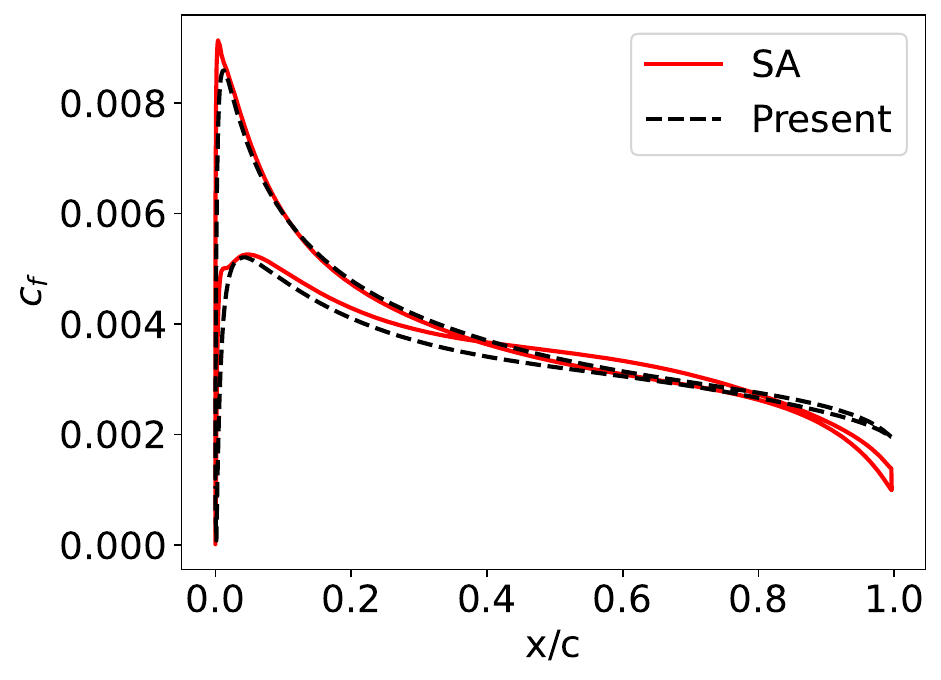} &
\includegraphics[width=0.47\textwidth]{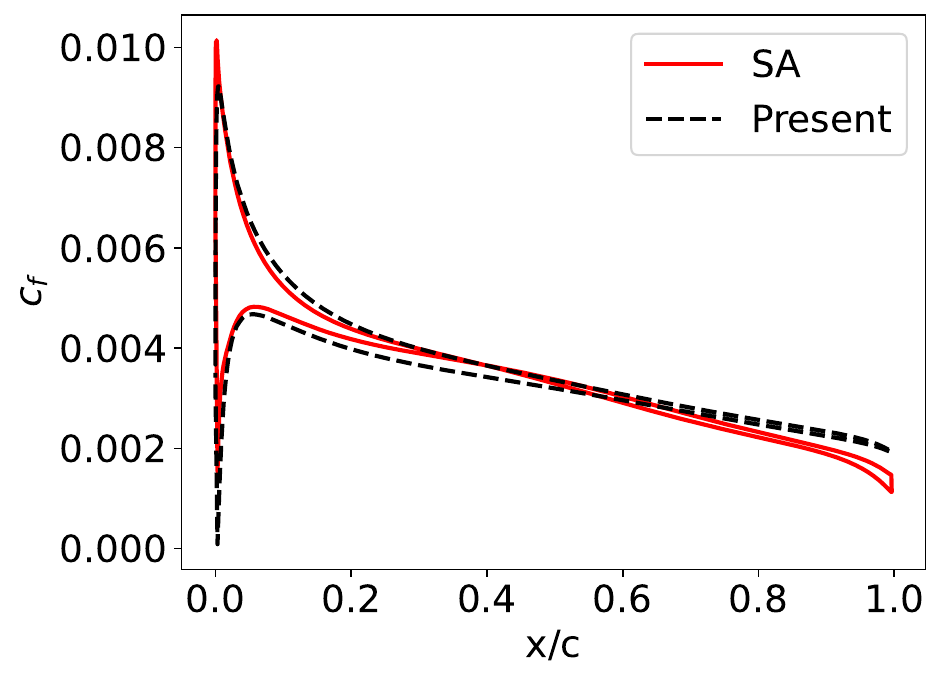} \\
\multicolumn{2}{@{}l@{}}{(e)} \\
\multicolumn{2}{c}{\includegraphics[width=0.72\textwidth]{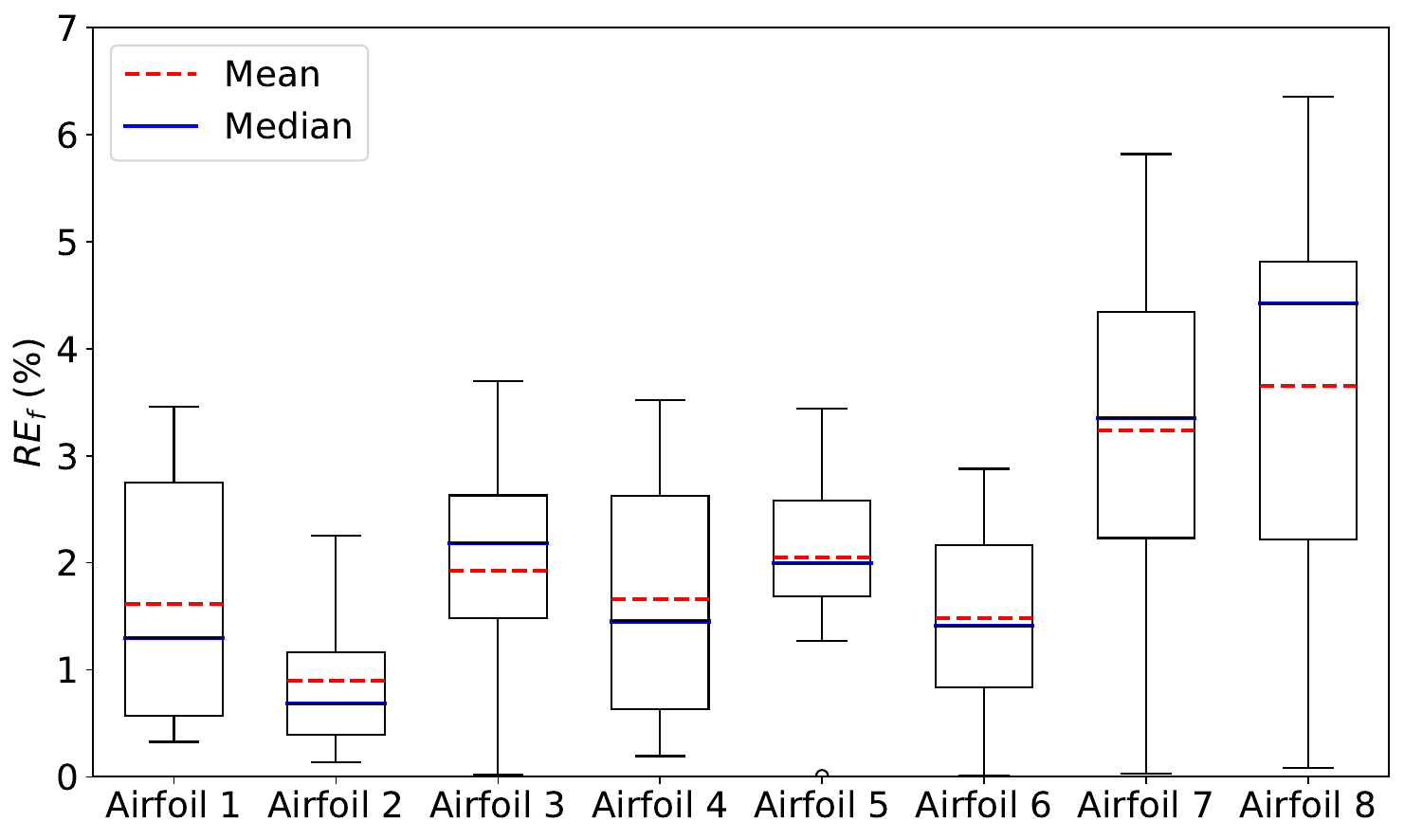}}
\end{tabular}
\caption{Generalization validation results for two-dimensional variable geometries. (a) The eight new airfoil geometries generated by CST parameter perturbations; (b--d) predictions for variable-geometry airfoil cases, where (b) corresponds to Airfoil 8; (e) box plot of integrated drag errors for the predictions on the eight airfoils.}
\label{fig:newairfoil}
\end{center}
\end{figure}

To further evaluate the generalization capability of the discovered expression chain to two-dimensional variable-geometry airfoils, the Pre\_newairfoil dataset from \linkedref{Table}{tab:datasets} is used for validation, and the skin friction distribution formula given by \eqref{eq:stage3} is adopted. This dataset takes the NACA0012 airfoil as the baseline. Random perturbations of $\pm30\%$ are applied to the shape function coefficients using the 12-level CST parameterization method, generating eight new airfoils (Airfoil 1 to Airfoil 8). Their geometric shapes are shown in \linkedrefpart{Figure}{fig:newairfoil}{(a)}. For each airfoil, ten operating conditions are randomly sampled from the freestream condition space, resulting in a total of 80 cases. The geometries of all these cases have not appeared in the training set, aiming to test the extrapolation capability of the expression chain to unseen geometries (shown in \linkedrefpart{Figures}{fig:newairfoil}{b--d}).

\linkedrefpart{Figure}{fig:newairfoil}{(e)} presents a box plot of the integrated drag errors for all 80 cases of the eight airfoils. The results show that for most perturbed airfoils (Airfoil 1 to Airfoil 6), the mean integrated drag error of the expression chain predictions is within 2\%. For airfoils with more pronounced geometric variations (Airfoil 7 and Airfoil 8), the error increases slightly but the mean value remains below 4\%, and the error for all airfoils is within 7\%. Overall, the expression chain exhibits good generalization capability for unseen geometries, with integrated drag errors all within the engineering acceptable range. This generalization capability benefits from the fact that the input features ($\Rex$, $\cp$, and $\Ma_x$) are all local flow parameters; together with the streamwise coordinate, they help the model accurately identify the flow development at each location, thereby effectively adapting to local shape variations of the airfoil. In scenarios requiring higher accuracy, the model can be further tuned using additional data and the existing theoretical formula framework to achieve more precise predictions.

\begin{figure}[!t]
\begin{center}
\begin{tabular}{@{}cc@{}}
\multicolumn{2}{@{}l@{}}{(a) $\Ma=0.36, Re=6.57\times10^6, \alpha=0.6^\circ$} \\
\includegraphics[height=0.26\textwidth]{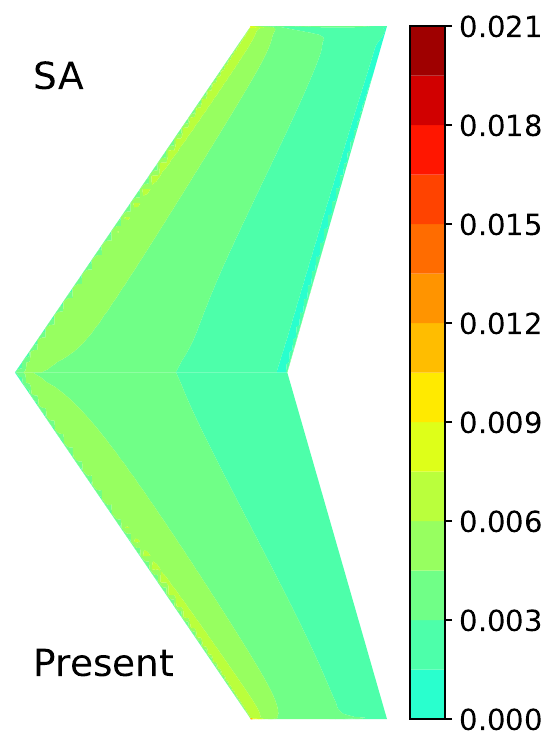} &
\includegraphics[width=0.36\textwidth]{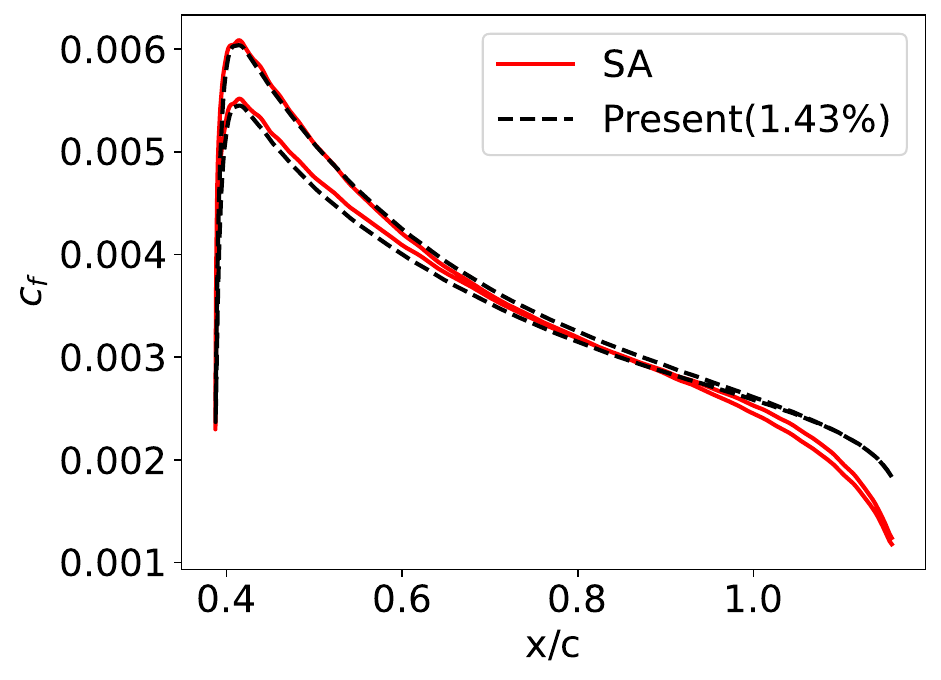} \\
\includegraphics[width=0.36\textwidth]{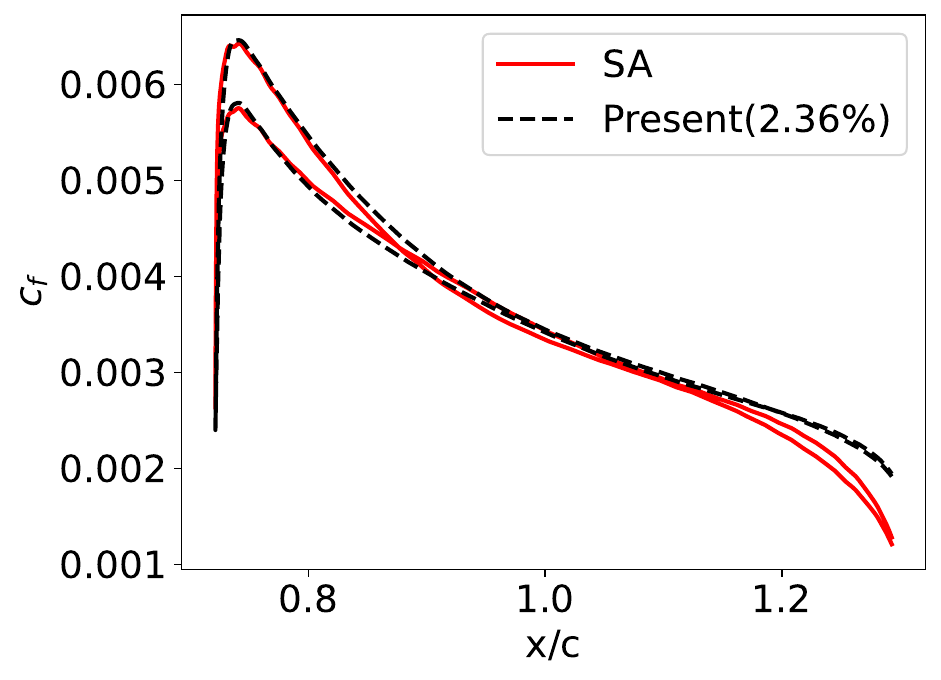} &
\includegraphics[width=0.36\textwidth]{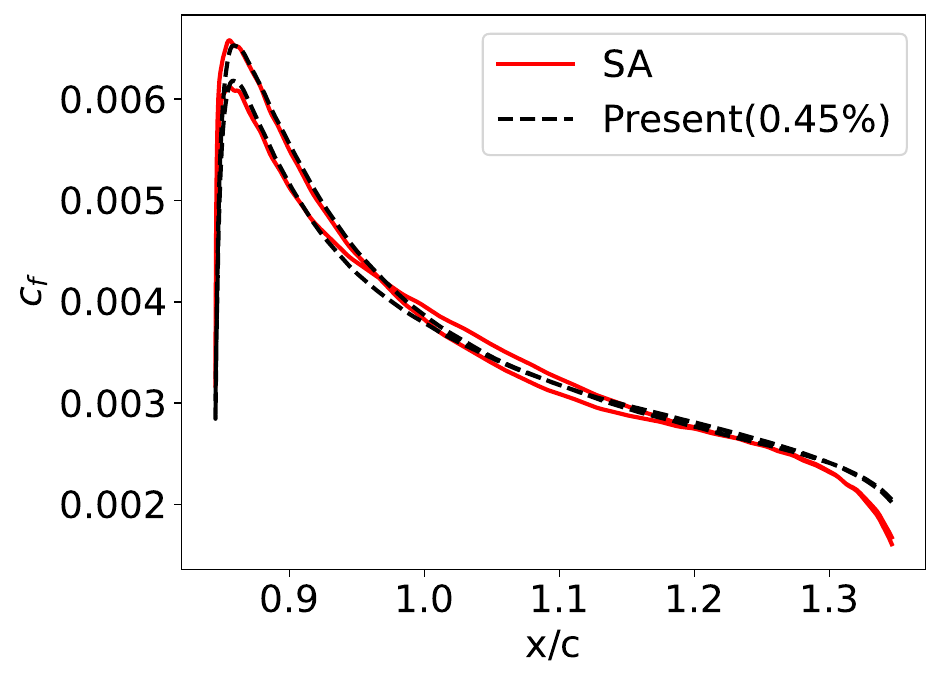} \\[6pt]
\multicolumn{2}{@{}l@{}}{(b) $\Ma=0.5, Re=6.0\times10^6, \alpha=3.74^\circ$} \\
\includegraphics[height=0.26\textwidth]{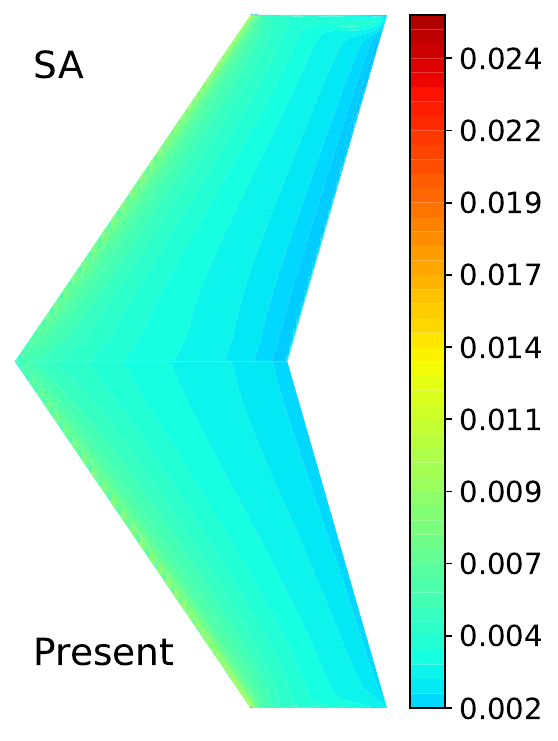} &
\includegraphics[width=0.36\textwidth]{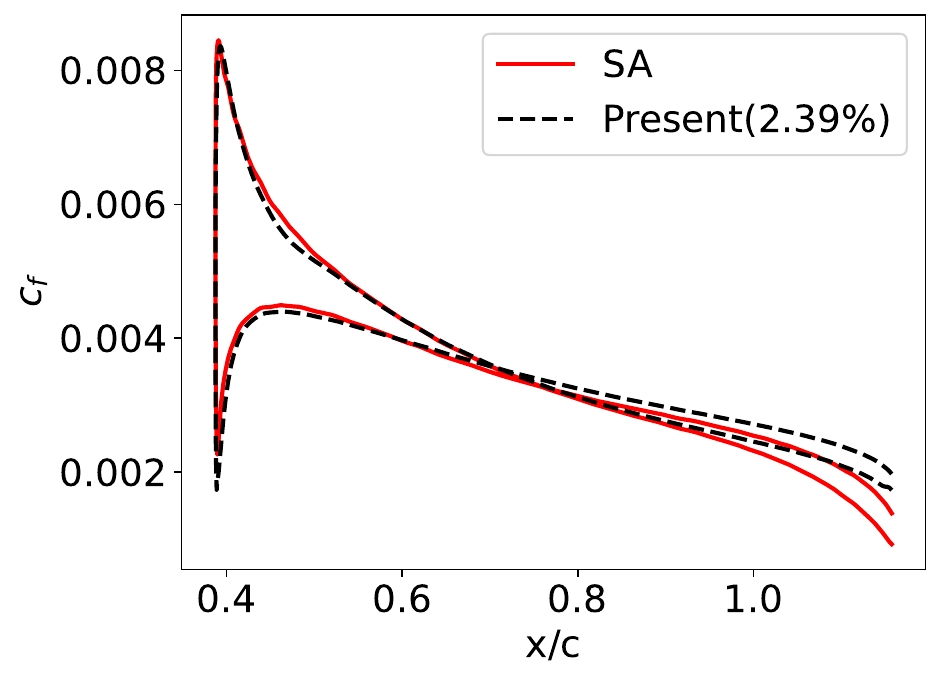} \\
\includegraphics[width=0.36\textwidth]{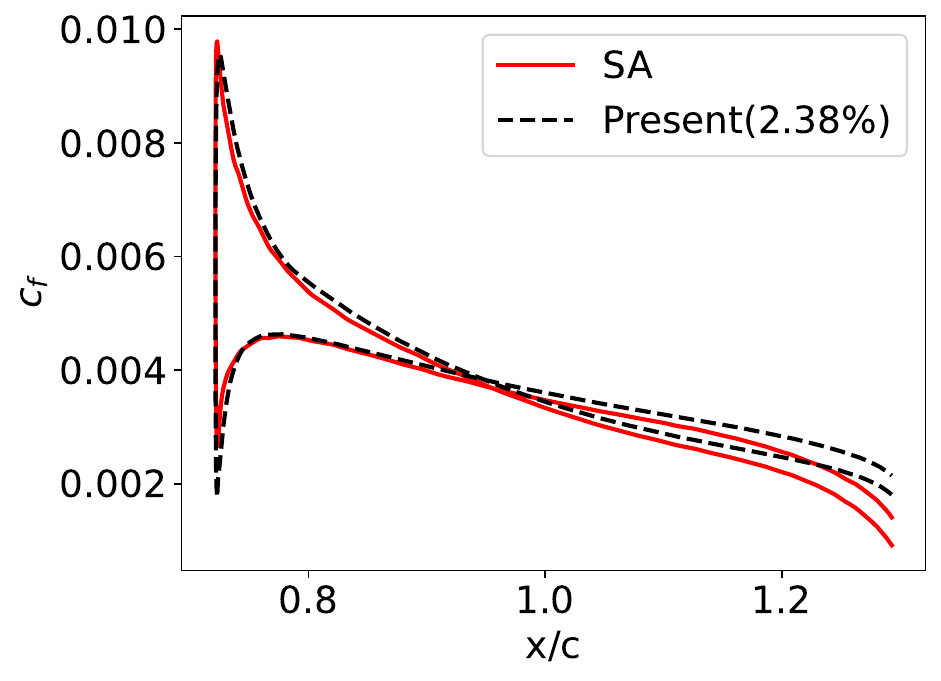} &
\includegraphics[width=0.36\textwidth]{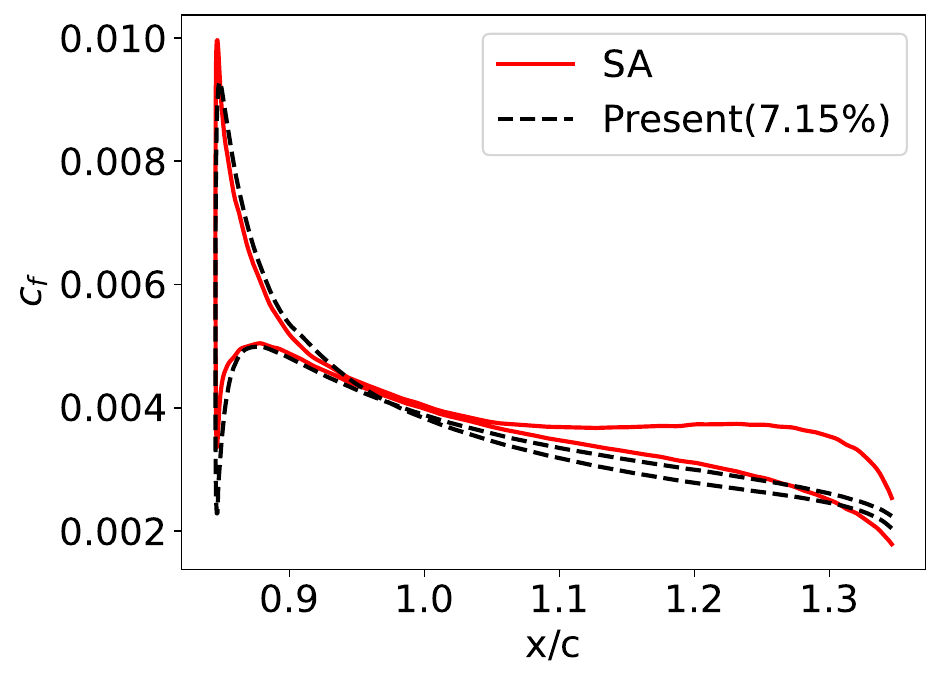}
\end{tabular}
\caption{Generalization validation results for a three-dimensional wing. Contour plot of the predicted skin friction distribution on the wing and comparison of skin friction at the 1/4, 1/2 and 3/4 span sections (top-right, bottom-left, bottom-right, respectively).}
\label{fig:wing}
\end{center}
\end{figure}

To further evaluate the generalization capability of the discovered expression chain to three-dimensional wing configurations with varying geometries, the Pre\_wing dataset from \linkedref{Table}{tab:datasets} is used for validation, and the skin friction distribution is computed using \eqref{eq:stage3}. This dataset consists of subsonic flows over variable-geometry three-dimensional wings. The wing under consideration has a taper ratio of 2, a span of 1.5 m, and a sweep angle of $30^\circ$. Its geometry is generated by extruding the NACA0012 airfoil, with shape variations achieved through CST parameter control and perturbations applied to the airfoil shape. Validation results are obtained under different freestream conditions and wing geometries, as shown in \linkedref{Figure}{fig:wing}. For case (b) ($\Ma=0.5$), away from the wing tip, the three-dimensional spanwise effect is weak and the two-dimensional formula remains applicable; near the wing tip, however, the spanwise effect becomes significant, leading to relatively large deviations in the trailing-edge prediction. For a finite-span wing, the downwash and cross-flow induced by the wingtip vortex thin the boundary layer, increase the velocity gradient, and consequently produce a local skin friction coefficient notably higher than that of a two-dimensional airfoil. Nevertheless, the mean relative errors of the integrated force for cases (a) and (b) are only 1.76\% and 2.15\%, respectively, demonstrating that the expression chain still possesses good generalization prediction capability for three-dimensional wings.

\section{Conclusions}
\label{sec:conclusions}

Addressing the demand for rapid skin friction prediction in aircraft design, this paper proposes a knowledge discovery method for skin friction drag that integrates physical knowledge with symbolic regression. This method demonstrates the capability of symbolic regression in fluid mechanics knowledge discovery and provides a paradigm for extracting physical laws from data and constructing interpretable models. The main conclusions are as follows:
\begin{list}{(\roman{enumi})}{%
  \usecounter{enumi}
  \setlength{\leftmargin}{2.6em}
  \setlength{\labelwidth}{1.8em}
  \setlength{\labelsep}{0.5em}
  \setlength{\itemsep}{0pt}
  \setlength{\parsep}{0pt}}
\item Construction of a white-box knowledge chain: using inviscid surface flow features as inputs, an analytical expression chain for the skin friction coefficient is successfully mined via symbolic regression. The expressions are concise in form, possess clear physical meaning, and exhibit a unified mathematical structure across different Mach number ranges, with only the scaling exponent adjusting according to the Mach number regime. This reveals the correction law of skin friction with varying flow parameters.
\item Physical interpretability and engineering value: the analytical formula chain provides aerodynamic engineers with an explicit drag model that can be directly embedded into design optimization frameworks, avoiding repetitive CFD computations. Moreover, its white-box nature facilitates understanding of the contribution mechanisms of different flow phenomena to skin friction.
\item Strong generalization capability: owing to its concise analytical form and clear physical meaning, the proposed expression chain demonstrates good extrapolation generalization capability for varying operating conditions and geometric configurations, including two-dimensional airfoils and three-dimensional wings.
\end{list}

This study still has several limitations that require further investigation. First, the current method is primarily validated for attached flow conditions. For transonic and high-angle-of-attack cases involving strong nonlinear flow features such as shock waves, whether symbolic regression can effectively capture the abrupt variation patterns of skin friction remains to be examined. Second, the adopted feature set is limited to inviscid surface flow parameters, with viscosity information only accounted for via the Reynolds number. Viscous information within the boundary layer, such as velocity profiles and turbulence intensity, has not yet been introduced, which may restrict the model's ability to characterize complex pressure gradient effects. Furthermore, the stage sequence and expression freezing strategy in the progressive discovery procedure have a certain influence on the final results; how to adaptively determine the optimal discovery path requires further exploration. Future work can extend the method to skin friction modeling for complex flows involving shock waves and separation, as well as explore deeper integration with other physical knowledge (e.g. boundary layer theory) to achieve higher-fidelity white-box modeling.

\par\vspace{2em}\noindent\textbf{Author contributions.} Mingkun Xia: Methodology, Validation, Visualization, Writing-original draft, Writing-review\&editing; Shule Zhao: Investigation, Methodology, Validation; Weiwei Zhang: Conceptualization, Funding acquisition, Investigation, Methodology, Project administration.

\par\medskip\noindent\textbf{Declaration of interests.} The authors report no conflict of interest.

\par\medskip\noindent\textbf{Data availability statement.} Data will be made available on request.

\par\medskip\noindent\textbf{Acknowledgements.} This research was supported by the National Natural Science Foundation of China (No.92152301, No. U2441211, No. U23B6009).

\appendix
\renewcommand{\theHsection}{appendix.\thesection}

\refstepcounter{section}
\section*{Appendix \thesection. Derivation of an Explicit Approximate Formula for the Skin Friction Coefficient in the Turbulent Boundary Layer}
\label{appA}

In a turbulent boundary layer, the velocity distribution law at the outer edge can be written as

\begin{equation}
  \frac{U_e}{u_\tau}
  =\frac{1}{\kappa}\ln\left(\frac{u_\tau\delta}{\nu}\right)+B+\frac{2\Pi}{\kappa},
  \label{eq:A1}
\end{equation}
where $U_e$ is the free-stream velocity at the boundary layer edge, $u_\tau=\sqrt{\tau_w/\rho}$ is the friction velocity, $\delta$ is the boundary layer thickness, $\nu$ is the kinematic viscosity, $\kappa=0.41$ is the von Karman constant, $B=5.0$ is the logarithmic law constant, and $\Pi=0.45$ is the Coles wake parameter.

Introducing the friction coefficient $\cf=\tau_w/(\rho U_e^2/2)$, we have $U_e/u_\tau=\sqrt{2/\cf}$ and $u_\tau\delta/\nu=Re_\delta\sqrt{\cf/2}$, where $Re_\delta$ is the Reynolds number based on boundary layer thickness. Substituting into \eqref{eq:A1} and inserting the constant values yields

\begin{equation}
  \sqrt{\frac{2}{\cf}}=2.44\ln\left(Re_\delta\sqrt{\frac{\cf}{2}}\right)+7.20.
  \label{eq:A2}
\end{equation}

In more general cases, the friction coefficient is often expressed in the following form:

\begin{equation}
  \frac{1}{\sqrt{\cf}}=b_1\ln \cf+b_2\ln Re+b_3,
  \label{eq:A3}
\end{equation}
where $b_1$, $b_2$ and $b_3$ are empirical constants, and $Re$ is the Reynolds number based on a characteristic length (the choice depends on the flow configuration; in this paper, the Reynolds number $\Rex$ based on streamwise length is adopted).

Equation \eqref{eq:A3} holds for flat plate conditions ($U_e=U_\infty$, pressure coefficient $\cp=0$). For more complex conditions, a pressure coefficient correction must be introduced. According to Bernoulli's principle (applicable to incompressible flow), $\cp=1-(U_e/U_\infty)^2$. Therefore, the pressure-corrected friction coefficient $\tilde{\cf}=\tau_w/(\frac{1}{2}\rho U_e^2)=\tau_w/(\frac{1}{2}\rho U_\infty^2)\cdot U_\infty^2/U_e^2=\cf/(1-\cp)$, with the pressure correction term being $1/(1-\cp)$. Then

\begin{equation}
  \frac{1}{\sqrt{\dfrac{\cf}{1-\cp}}}
  =b_1\ln\left(\frac{\cf}{1-\cp}\right)+b_2\ln(\Rex)+b_3.
  \label{eq:A4}
\end{equation}

To convert the implicit form of \eqref{eq:A4} into an explicit form, we refer to the transformation method of \citet{shi1988explicit}, introducing the variables

\begin{equation}
  z=\frac{1}{2b_1}\sqrt{\frac{1-\cp}{\cf}},
  \qquad
  a=\frac{\Rex^{b_2/(2b_1)}
  \exp\left\{\dfrac{b_3-b_1\ln(1-\cp)}{2b_1}\right\}}
  {\dfrac{2b_1}{\sqrt{1-\cp}}}.
  \label{eq:A5}
\end{equation}

Equation \eqref{eq:A4} then simplifies to the simple form $z\mathrm{e}^{z}=a$, where the relationship between $z$ and $a$ can be obtained by solving the implicit equation. For engineering convenience, the following piecewise explicit approximation is given for $a\in[46,1.6\times10^7]$: $z=b_4(\ln a)^{b_5}$. This range covers the typical Reynolds number regimes of engineering turbulent boundary layer flows, from the low-Reynolds-number transition region to the high-Reynolds-number fully turbulent region. Substituting back into \eqref{eq:A5} gives

\begin{equation}
  \frac{1}{\dfrac{2b_1}{\sqrt{1-\cp}}\sqrt{\cf}}
  =b_4\left[
  \frac{b_2}{2b_1}\ln(\Rex)
  +\frac{b_3-b_1\ln(1-\cp)}{2b_1}
  +\ln\left(\frac{\sqrt{1-\cp}}{2b_1}\right)
  \right]^{b_5}.
  \label{eq:A6}
\end{equation}

Further simplification leads to

\begin{equation}
  c_f = \frac{(1 - C_p) \frac{\left(\frac{2b_1}{b_2}\right)^{2b_5}}{4(b_1)^2(b_4)^2}}{\left(\ln(Re_x) + \frac{b_3}{b_2} - \frac{2b_1}{b_2} \ln(2b_1)\right)^{2b_5}}.
  \label{eq:A7}
\end{equation}
It is observed that this form is highly consistent in structure with the explicit approximate formula $\cf=(1-\cp)/(\log\Rex)^{3.18}$ discovered by our symbolic regression, validating the theoretical rationality of the symbolic regression results.

The basic expression discovered by symbolic regression in stage one is $\cf^{(0)}(Re,L)=1/(\log(Re\cdot L))^n$, where the exponent $n\approx3.18$. Starting from classical flat-plate turbulent friction formulas, we back-calculate the equivalent exponent via coefficient normalization and cross-validate with the symbolic regression results. Classical flat-plate turbulent boundary layer friction coefficient formulas are

\begin{align}
  \mbox{Prandtl--Schlichting (1932):}\quad
  \cf &=0.455(\log\Rex)^{-2.58},
  \label{eq:ps}\\
  \mbox{Wieghardt (1955):}\quad
  \cf &=0.52(\log\Rex)^{-2.685}.
  \label{eq:wieghardt}
\end{align}
Setting the two equations equal, i.e. $0.455(\log\Rex)^{-2.58}=0.52(\log\Rex)^{-2.685}$, an averaged Reynolds number can be solved. Using this Reynolds number as a reference and normalizing the coefficient of the formula to 1, i.e. $0.455(\log\Rex)^{-2.58}=(\log\Rex)^{-n}$, we obtain $n\approx3.199$, which is very close to the 3.18 obtained from symbolic regression.

Starting from the velocity distribution law for turbulent boundary layers, this appendix derives the implicit expression for the skin friction coefficient. After introducing pressure correction, the standard form $z\mathrm{e}^z=a$ is obtained through variable transformation. An explicit approximation and the validation of the exponent scaling are provided. The final form is highly consistent with the symbolic regression results, demonstrating the rationality and practicality of the explicit approximate formula.

\begingroup
\setlength{\bibsep}{5pt}
\bibliographystyle{unsrtnat}
\bibliography{ref}
\endgroup

\end{document}